\documentclass[final]{siamart1116}

\usepackage{algorithm}
\usepackage{algpseudocode}
\usepackage{amsmath} 
\usepackage{amsfonts}

\newcommand{\TheTitle}{CaverDock: A Novel Method for the Fast Analysis of Ligand Transport}
\newcommand{\RunningTitle}{A Novel Method for the Fast Analysis of Ligand Transport}
\newcommand{\TheAuthors}{J Filipovi\v{c} et al.}

\def\eg{e.\,g.}
\def\ie{i.\,e.}

\newcommand{\Tau}{\mathrm{T}}
\newcommand{\angstrom}{\textup{\AA}}

\theoremstyle{definition}
\newtheorem{defn}{Definition}[section]

\DeclareMathOperator{\dst}{dst}
\DeclareMathOperator{\dis}{distance}
\newcommand{\norm}[1]{\left\lVert#1\right\rVert}
\newcommand{\Break}{\State \textbf{break} }

\headers{\RunningTitle}{\TheAuthors}

\title{\TheTitle}
\author{
  Ji\v{r}\'{i} Filipovi\v{c}\thanks{Institute of Computer Science, Masaryk University
    (\email{fila@mail.muni.cz}, \email{408420@mail.muni.cz}, \email{ludek@ics.muni.cz}).}
  \and
  Ond\v{r}ej V\'{a}vra\thanks{Department of Experimental Biology and RECETOX, Faculty of Science, Masaryk University
     (\email{393643@mail.muni.cz}, \email{222755@mail.muni.cz}, \email{smarques@mail.muni.cz}, \email{brezovsky@mail.muni.cz}, \email{jiri@chemi.muni.cz}).}
  \and
  Jan Plh\'{a}k\footnotemark[1]
  \and
  David Bedn\'{a}\v{r}\footnotemark[2]
  \and
  S\'{e}rgio M. Marques\footnotemark[2]
  \and
  Jan Brezovsk\'{y}\footnotemark[2]
  \and
  Lud\v{e}k Matyska\footnotemark[1]
  \and
  Ji\v{r}\'{i} Damborsk\'{y}\footnotemark[2]
}

\begin{document}

\maketitle

\begin{abstract}

Here we present a novel method for the analysis of transport processes in proteins and its implementation called CaverDock. Our method is based on a modified molecular docking algorithm. It iteratively places the ligand along the access tunnel in such a way that the ligand movement is contiguous and the energy is minimized. The result of CaverDock calculation is a ligand trajectory and an energy profile of transport process. CaverDock uses the modified docking program Autodock Vina for molecular docking and implements a parallel heuristic algorithm for searching the space of possible trajectories.
Our method lies in between the geometrical approaches and molecular dynamics simulations. Contrary to the geometrical methods, it provides an evaluation of chemical forces. However, it is far less computationally demanding and easier to set up compared to molecular dynamics simulations. CaverDock will find a broad use in the fields of computational enzymology, drug design and protein engineering. The software is available free of charge to the academic users at https://loschmidt.chemi.muni.cz/caverdock/.

\end{abstract}



\section{Introduction}

Understanding protein-ligand interactions is of great importance in many fundamental biochemical processes as well as in various applications. For example, to study a ligand that may inhibit protein function allowing a virus to attack a cell or to design inhibitors blocking tunnels and channels as a new paradigm in drug design~\cite{marques2017enzyme}. The ligand interacts with protein in its \textit{active} or \textit{binding site} -- the functional site of a protein. Simulation of ligand binding (entering the active site and forming a stable complex) and unbinding (release of a ligand from a stable complex) helps in many practical applications. It allows to search for ligands which are more likely to bind to a particular protein; modify a ligand to bind faster or with higher affinity or modify the protein to ease or disallow the ligand binding. Many proteins have binding sites buried inside their cores, which implies that a ligand must traverse through a \textit{tunnel} or a \textit{channel}\footnote{In the following text, we will speak about tunnels for simplicity. However, the mechanisms of a ligand passing through a channel are the same as for a tunnel.} before it can bind to the functional site in the protein. In such cases, we need to analyze whether the ligand is likely to pass through the tunnel or channel into the protein core.

All chemical systems, such as the proteins interacting with ligands, follow the second law of thermodynamics: they tend to minimize their potential energy. In practice, the most probable \textit{conformation} of the molecules (\ie{} spatial position of their atoms) is the one with the lowest potential energy. However, it is also possible for molecules to make a transition from one local minimum to another, depending on system temperature and height of energetic barrier -- the smaller energetic barrier, the more probable is the transition. 
In the molecular modeling methods, the function estimating the potential energy for a given conformation is called a \textit{force field}. The analysis of the potential energy given by the force field allows us to compute the probability of some conformation to appear in the real-world chemical system. It allows to predict whether or how fast some chemical process (such as a ligand passing through a tunnel) can occur at a given temperature.

To study the ligand binding or unbinding, we need to evaluate the potential energy of the ligand passing from protein surface through the tunnel into the active site or vice versa. The ligand binds in an active site if there is a strong local energetic minimum and it passes through the tunnel if there is no significant energy barrier along the way (the gradient of potential energy is more or less decreasing from tunnel entrance to its binding site). When a tunnel contains some strong repulsive barrier, the ligand is likely not to pass through the tunnel. Note that the energetic profile of the tunnel is unique with respect to the ligand, as it reflects the specific ligand-protein interactions occurring during the ligand passage.

The ligand binding to a protein's active site or binding site is usually computed by \textit{molecular docking}. A molecular docking algorithm traverses the conformation space of the protein-ligand complex and searches for energetic minima~\cite{trott2010autodock, ewing2001dock, thomsen2006moldock, morris2009autodock}. The result of the molecular docking is the structure of the protein-ligand complexes together with an estimation of the respective free energy of binding. Thus, the users can learn which ligand binds with the lowest energy or study the orientation of a ligand in a protein active site. However, the molecular docking computes only the lowest energy positions of a ligand within some region of a protein, and thus it is not suitable for the study of the ligand transport through a protein tunnel.

In this paper, we present a novel method for computation of the binding free energy variation and the ligand \textit{trajectory} (movement of the ligand's atoms along the tunnel). It allows to study the processes of ligand binding and unbinding through the tunnels of any protein. Our method is based on a molecular docking algorithm -- it iteratively docks the ligand along the previously calculated tunnel and at each point it evaluates its binding free energy. Our docking works with a novel hybrid force-field. It uses a combination of the chemical force-field from AutoDock Vina~\cite{trott2010autodock} to compute the binding free energy of the protein-ligand complex and a newly developed constraints force-field, which restricts the ligand positions to specified part of the tunnel and to vicinity of defined conformation. With the constraints, a contiguous ligand trajectory with arbitrary step size (the maximal change in ligand's atoms position between two consecutive conformations) can be generated. Thus, the position of the ligand within the tunnel can be constrained to a defined area. Since there may exist many possible paths through a given tunnel, the paths are searched using a heuristic algorithm with backtracking. Our method is implemented in the user-friendly software tool CaverDock, which uses parallel architecture to maximize the performance of the ligand transit computation (from minutes to a few hours using a desktop computer).

This paper focuses primarily on the computer-science topics: it introduces our method and its implementation. It also presents basic evaluation, which illustrates that the produced trajectories and energetic profiles can be obtained in a reasonable time. The paper targeting the CaverDock user community, focused on the biochemical topics (setting the calculation and interpretation of results, evaluation and benchmarking CaverDock with realistic use cases on many protein-ligand pairs) is prepared in parallel with this paper~\cite{vavra2018caverdock, pinto2018fast}.

The rest of the paper is organized as follows. Section~\ref{sect:relate_work} summarizes the work related to our paper and describes the difference between our method and state-of-the-art. The high-level overview of our method is given in Section~\ref{sect:method_overview}. Following three sections discuss the method in detail: Section~\ref{sect:discretization} introduces the algorithm for tunnel discretization, Section~\ref{sect:constrained_docking} describes our modifications of docking algorithm using constraints for ligand position and Section~\ref{sect:trajectory_search} discuss the searching of trajectory space. The evaluation of our implementation is given in Section~\ref{sect:evaluation}. We conclude and sketch future work in Section~\ref{sect:conclusion}.

\section{Related Work}
\label{sect:relate_work}
A very fast approximation of biomolecules represents the atoms as solid macro-world objects. The transport process of a ligand through a tunnel in a protein is then studied analogously to a macro-world objects' mechanics: the molecular shape is formed by spheres representing the atoms (which may be connected through flexible joints), neglecting any chemical forces, such as electrostatics, hydrogen bonds or solvation effects.

The majority of the geometry-based approaches analyze the tunnel only, without generating a ligand trajectory~\cite{chovancova2012caver, sehnal2013mole, yaffe2008molaxis}. Those software tools take a protein or multiple conformations of the protein as the input and generate the geometry of the tunnel. The ability to transport a ligand is then judged based on tunnel geometry. A comprehensive study of different geometrical methods can be found in~\cite{brezovsky2013software, krone2016visual}.

A different approach to the geometric analysis is taken in MoMA-LigPath~\cite{devaurs2013moma}. The ligand transport is studied here by an algorithm inspired by robotic motion planning. The protein and ligand are understood as mechanical objects, which are partially flexible as they may change the dihedral angles. The algorithm searches for a ligand trajectory from the active site to the tunnel entrance by moving the ligand and the flexible parts of the receptor. Such algorithm allows to detect parts of the receptor which need to be moved to allow the ligand to pass through the tunnel. However, it does not use a chemical force field, so there is no quantitative information showing how difficult is for the ligand to pass the tunnel due to chemical interactions (attractions and repulsions). Moreover, the induced movement of the ligand and the flexible parts may be unrealistic, as with the chemical forces different movements may be preferred.

The molecular dynamics (MD) uses an empirical force field to model the physical properties of the atoms and their interactions in time. There are many well-established software tools for MD, such as Amber~\cite{salomon2013overview} or Gromacs~\cite{abraham2015gromacs}. However, it is not practical to model the transportation of a ligand through a tunnel with classical MD, as the simulation time is often extremely long.
Therefore, various modifications of MD are used to speed-up the process.

The metadynamics is an enhanced sampling technique which introduces biases in the form of repulsion energy on the already visited parts of the conformational space, such as the conformations of a molecule or the positions of a ligand within a tunnel~\cite{branduardi2007from}. The bias is computed according to the simulation state defined in term of collective variables (a small number of variables describing the simulation space). The metadynamics can be used to pass a ligand through a tunnel in a protein~\cite{tiwary2015kinetics} and evaluate the thermodynamics and kinetics of the process. It explores simulation states much faster than MD, however, comparing to the geometrical approaches, it is still much more computationally demanding. Moreover, an expert user has to setup the metadynamics computation properly, as an incorrect definition of the collective variables may lead to inefficient biassing.

Another technique based on MD allowing to simulate transportation through a tunnel is the steered MD~\cite{li2012steered}. With steered-MD, the external force is applied to a ligand such that it is pulled from or to the tunnel. The technique is, similarly to metadynamics, more computationally demanding than the geometrical methods. An expert user has to set up how the external force is applied, otherwise, there can be a false bottleneck observed (\eg{} when a ligand is pulled in the wrong direction against the protein backbone).

The molecular docking has been developed for evaluation of the ligand binding free energy in the protein active site. It performs the search of many protein-ligand conformations and returns the ones in significant energetic minima. Thus, it is possible to study if the ligand binds preferably in the active site, or compare the minima of different ligands (\ie{} to identify which ligands are more likely to interact with the protein). Many docking software tools are well-established and widely-used in the scientific community~\cite{trott2010autodock, ewing2001dock, thomsen2006moldock, morris2009autodock}. Molecular docking is not suitable to be directly used for analysis of ligand transport, as it samples conformation space coarsely to find significant minima, but does not search the ligand path into the minima. However, it can be used as a basis for docking-based tools.

Similarly to our tool, molecular docking is used to analyze the transport process in SLITHER~\cite{lee2009slither}. However, the SLITHER does not employ constrained docking. Instead, the movement of the ligand is induced by the biasing force at the already visited positions. Therefore, there is no mechanism to ensure the ligand movement is contiguous or at least provides fine-grained sampling of the trajectory in the tunnel -- it may jump over bottlenecks without sampling the energy barriers or even jump into a different tunnel. Moreover, there is no sophisticated tunnel geometry analysis and the ligand is moved along the $y$-axis only. Therefore, SLITHER cannot be reliably used for highly curved tunnels, \eg{} U-shaped. 

\section{Method Overview}
\label{sect:method_overview}
In this section, we describe the basic concept of our method. The more detailed discussion will be given in the following sections. The method is based on a driven step-by-step movement of the ligand through the tunnel.
We first discretize the tunnel into a set of discs, so the ligand movement through the tunnel can be driven, \ie{} it is possible to define a ligand position in the tunnel and thus also movement "forward" and "backward" in the tunnel. After the discretization, the ligand is iteratively docked into consecutive positions along the tunnel, allowing to compute binding or unbinding trajectory.
 
\subsection{Tunnel Discretization}

To drive the ligand movement in the tunnel, we need to restrict the space where the ligand can be placed in each step. We use the tunnel geometry approximated by a sequence of spheres as the input. Such sequence can be obtained from Caver~\cite{chovancova2012caver} or a similar tool. The sequence of spheres is then transformed into a sequence of $n$ disks $\theta_1, \dots, \theta_n$. We create the disks by cutting the tunnel into slices of an upper-bound thickness.
The path of the ligand through the tunnel can be defined as the iterative placement of the selected ligand's atom to consecutive disks. Note that an arbitrary atom of the ligand can be selected, but it must be the same for the whole tunnel trajectory.

\subsection{Constrained Docking}
\label{sect:oberview_constraint}

\begin{figure}[t]
\centering
\includegraphics[width=.8\hsize]{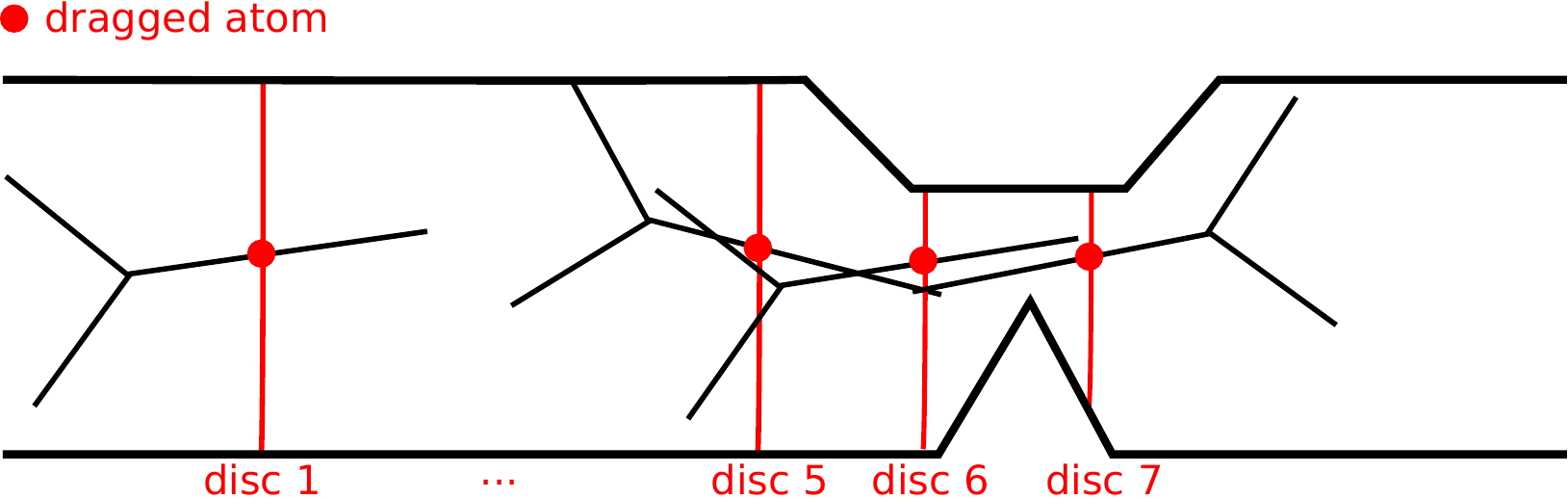}
\caption{Schematic 2D view of traversing tunnel, where the selected ligand's atom is placed onto consecutive disks. As no contiguous movement of the ligand is required, the ligand flips between disks $\theta_6$ and $\theta_7$, thus the small geometrical bottleneck between those disks is not detected.}
\label{fig:lower-bound}
\end{figure}

The ligand conformation $\lambda$ is defined by the Cartesian position of its atoms: $\lambda = \{a_i\}_{i=1}^m$. Having a discretization of our tunnel, we can select an atom of the ligand $a_c \in \lambda$, which is placed onto any position of the selected disk $\theta$: 
\begin{equation}
a_c \cap \theta = a_c
\label{eq:position}
\end{equation}
We say the ligand is docked onto the disc when its atom $a_c$ lies onto the disc. By placing the atom $a_c$ onto consecutive discs $\theta_1, \dots, \theta_n$, we force the ligand to move through the tunnel. Such ligand trajectory samples the tunnel without large gaps (\ie{} the ligand cannot overcome very narrow bottlenecks or even jump to different tunnel), however, the trajectory is not contiguous (the ligand can \eg{} rotate freely). We use this non-contiguous trajectory to compute the \textit{lower-bound} energy profile of the ligand transport. The example of such trajectory is depicted in Figure~\ref{fig:lower-bound}. As we can see, atom $a_c$ is stuck to the disk and by moving through the tunnel, the sampling of the transport process is obtained. However, the ligand may perform non-contiguous movement: it flips between disc $\theta_6$ and $\theta_7$. 

\begin{figure}[t]
\centering
\includegraphics[width=.5\hsize]{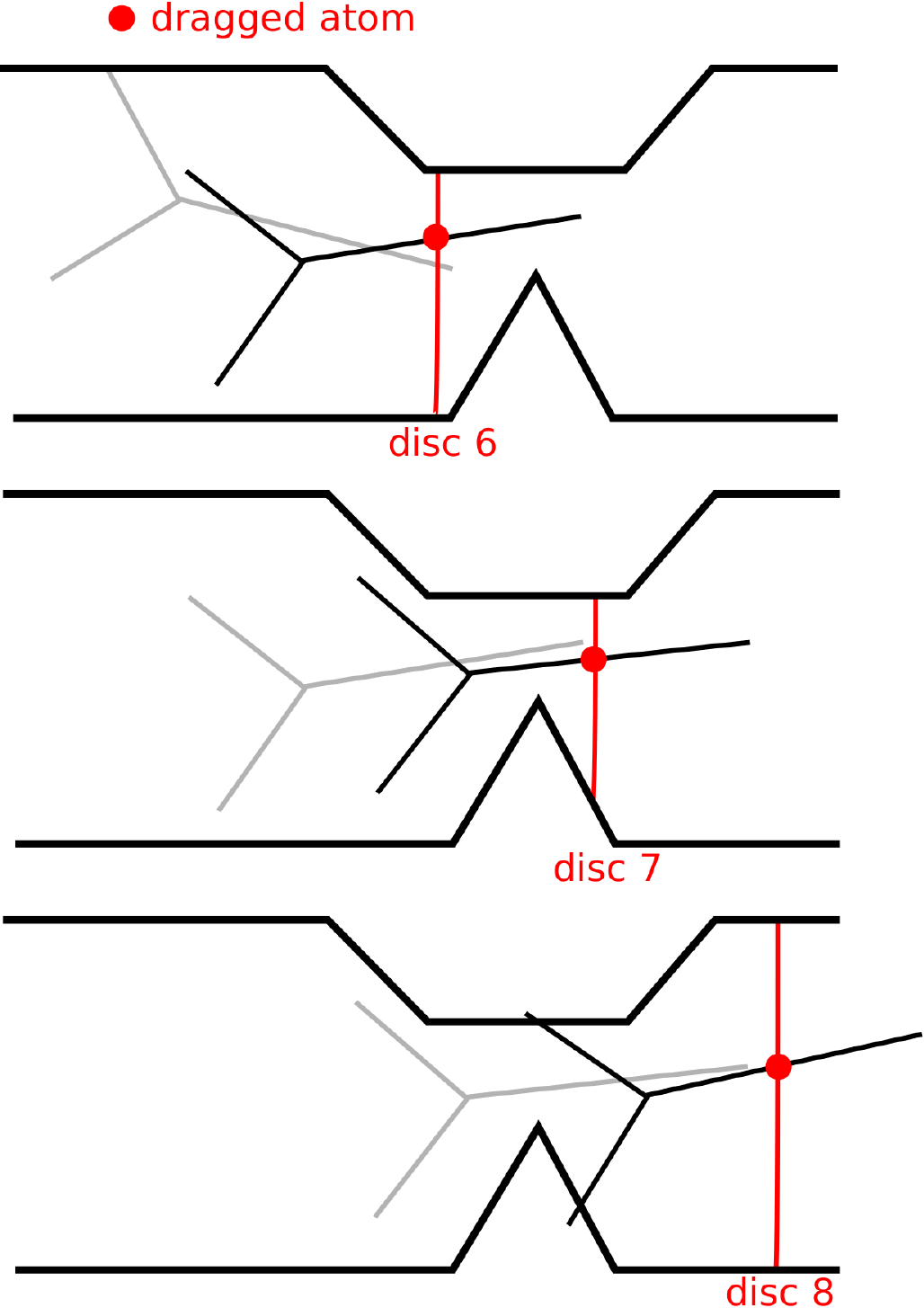}
\caption{Schematic 2D view of a ligand traversing a tunnel. The ligand is depicted in black, its previous position used as a pattern is shown in grey. Restricting the movement of atoms causes the geometrical bottleneck between $\theta_6$ and $\theta_7$ to be detected when the ligand passes from $\theta_7$ to $\theta_8$, as can be seen in the last figure.}
\label{fig:bottleneck}
\end{figure}

The contiguous trajectory can be computed by restricting the movement of each atom by constant $\delta$. When a new ligand conformation $\lambda_{i+1}$ is generated, the distance of each atom from its previous position in $\lambda_i$ is upper-bound:
\begin{equation}
\forall j \in [1,m] \Rightarrow |a_j - b_j| < \delta
\label{eq:pattern}
\end{equation}
where $a_j \in \lambda_i, b_j \in \lambda_{i+1}$.
We say $\lambda_{i}$ is the pattern constraining the position of $\lambda_{i+1}$, formally: $\lambda_{i+1} \in \Delta \lambda_i$, when Eq.~\ref{eq:pattern} holds. 

We can use the pattern constraint when a new position of the ligand is generated. Let $\lambda^i$ represents the ligand conformation docked onto disc $\theta_i$. When the ligand position $\lambda^{i+1}$ on the disc $\theta_{i+1}$ is searched, the pattern constraint ensures that $\lambda^{i+1} \in \Delta \lambda^i$, thus, transition between discs is contiguous (upper-bound by $\delta$). Note that the discs must be generated such that the distance between the discs must be lower than $\delta$. 
The example of using pattern constraint is depicted in Figure~\ref{fig:bottleneck}. As we can see, the pattern disallows the ligand to flip (as exemplified by the movement from disc 6 to disc 7 in Figure~\ref{fig:lower-bound}), and the small geometrical bottleneck is detected.

\subsection{Trajectory Search}

The contiguous trajectory can be obtained by iterative docking onto the disks with restricted changes in the position of all atoms by a pattern constraint. However, we want to allow the ligand to optimize its position at each disc to find a local energetic minimum. This minimum may be unreachable after one step when the ligand movement is restricted by a pattern. Thus, we search for the ligand trajectory, where multiple conformations may be docked onto the same disc. More precisely, having the conformation $\lambda_j^i$ at disc $\theta_i$, we search for conformation $\lambda_{j+1}^i \in \Delta \lambda_j^i, \lambda_{j+2}^i \in \Delta \lambda_{j+1}^i, \dots$ until the energy of the new conformations is improved. We call these steps the optimization steps, as they allow the ligand to find a low-energy position on the disc, which may not be feasible immediately after the transition from $\theta_{i-1}$ to $\theta_i$.

\begin{figure}[t]
\centering
\includegraphics[width=.8\hsize]{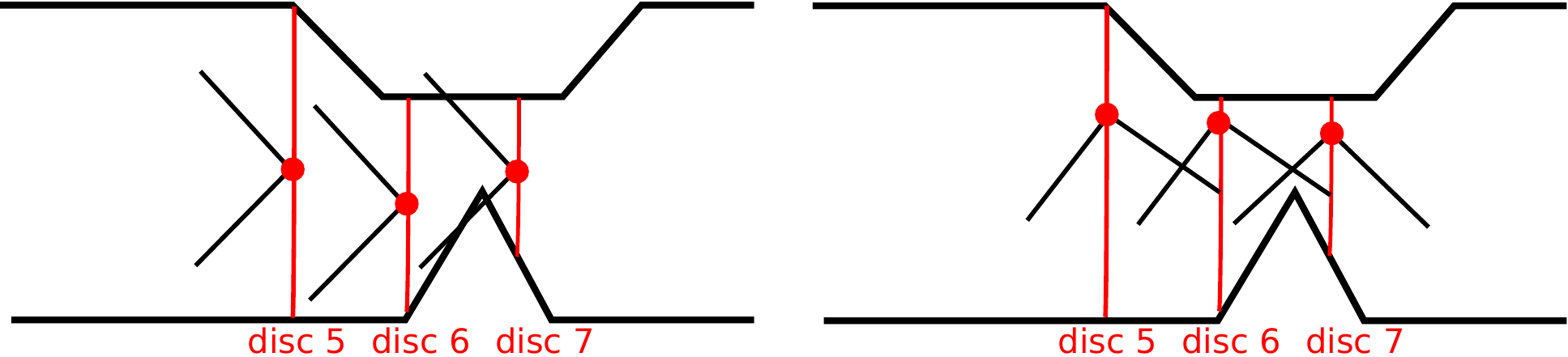}
\caption{Schematic 2D view of a ligand traversing a tunnel, where the ligand is stuck at a bottleneck (left figure). When a different orientation of the ligand is selected, the ligand may pass without reaching the barrier (right figure).}
\label{fig:backtracking}
\end{figure}

The ligand movement described above prefers the transition where the ligand follows the strongest energy gradient locally between following steps. Although this scenario is the most probable in real-world systems, the ligand may occasionally make a transition to some different conformation, which may allow it to pass the energy barrier with lower energy. Consider the case depicted in Figure~\ref{fig:backtracking}. Depending on its orientation, the ligand may or may not get through the tunnel bottleneck. Thus we need to search multiple variants of the ligand trajectory.

The number of contiguous trajectories may be very high -- the transition to a new disk may change the ligand position, orientation, and conformation (relative position of the atoms within the ligand). 
The exhaustive search of possible trajectories is not feasible due to the time required to dock ligands with constraints (typically hundreds of milliseconds). Thus, we have introduced a simple heuristic. We move the ligand only in one direction in the tunnel (\eg{} from the binding site to the protein surface). When the binding free energy of $\lambda_j^i$ is significantly higher than the binding free energy of some known conformation $\lambda_{low}^i$ (\ie{} obtained during lower-bound trajectory computation), we set $\lambda_j^i = \lambda_{low}^i$, and search the trajectory moving the ligand backwards to previous disks $\theta_{i-1}, \theta_{i-2}, \dots$. The backtracking ends after the forward and backward trajectories converge, or after the beginning of the tunnel is reached. Note that the resulting trajectory still follows only one direction. When the backtrack is used the trajectory is reversed and integrated into a forward trajectory. 

The situation when the backtracking trajectory converges with a forward trajectory (\ie{} $\lambda_{backtrack}^i \in \Delta \lambda_{forward}^i$) allows to join both trajectories. The optimization of the ligand position moves it to a minimum at the current disc which allows convergence in many cases. However, the ligand may need to overcome some energetic barrier to converge. Therefore, we use also an explicit convergence process: a weak force is applied to the ligand in the backtrack trajectory in order to pull its position to the vicinity of the ligand in the forward trajectory.

\subsection{CaverDock Workflow}
From the user's perspective, CaverDock is a com\-mand-line tool taking the molecules' structures and the tunnel geometry as input and producing the trajectory of the molecule and energetic profile as output. The CaverDock workflow consists of multiple steps:
\begin{enumerate}
  \item gather the input data (ligand in pdb or mol2 format, protein in pdb format), which can be obtained from experiments, downloaded from PDB~\footnote{https://www.wwpdb.org/} etc.
  \item convert the input data into PDBQT format using AutoDock Tools~\cite{morris2009autodock}
  \item identification and selection of a tunnel within the protein using Caver~\cite{chovancova2012caver}
  \item discretization of the tunnel exported from Caver by the CaverDock script \texttt{discretizer.py}
  \item (optional) setting the flexibility of selected side-chain residues by AutoDock Tools
  \item computing a box around the tunnel and the flexible residues either manually or using the CaverDock script \texttt{prepareconf.py}
  \item execution of CaverDock to search for the ligand trajectory
  \item analyze CaverDock trajectory and the energetic profile, and optionally identify new side chains which should be flexible and return to step 5
\end{enumerate}
CaverDock's script \texttt{flexibilize.py} allows to automatically search for flexible re\-si\-du\-es. The script first runs CaverDock with the rigid receptor, and then iteratively runs CaverDock allowing the flexibility on the side-chains which have formed the bottlenecks in the previous iteration.

\section{Tunnel Discretization}
\label{sect:discretization}

In this section we describe our requirements on the tunnel discretization in detail and the important parts of the algorithm performing the discretization.

\subsection{Tunnel Discretization Requirements}
\label{sect:tun_req}
As the first step in the CaverDock workflow, the tunnel must be discretized in discs, which will restrict the ligand's position in every docking. We use a geometric representation of the tunnel from Caver~\cite{chovancova2012caver}. It approximates the tunnel as a sequence of spheres $\Tau = \{S_i\}_{i=1}^k$, where the following holds:
\begin{equation}
  \begin{split}
    S_i \bigcap S_{i+1} \neq \emptyset; 1 \leq i < k \\
    S_i \not\subseteq S_j; \forall i \neq j
  \end{split}
\end{equation}
Moreover, the tunnel $\Tau$ never intersects itself, \ie{} it is topologically equivalent to a cylinder.

Recall that the movement of the ligand is determined by the placement of its atom (the drag atom $a_c$) to the discs. 
Thus, we need to transform the tunnel $\Tau$ to a sequence of discs.

\begin{defn}
The cut $\theta$ of tunnel $\Tau$ is a disc in the three-dimmensional space, which is defined by a triple $\theta = (A, u, r)$, where $A\in \mathbf{R}^3$ is a centre, $ u\in \mathbf{R}^3$ is a normal and $r>0$ is a radius. The $\Tau \cap \theta$ must be a continuous set and $\exists \delta > 0$ such that  $\forall \varepsilon > 0,  \varepsilon < \delta$ holds $(A, u, r + \varepsilon) \cap \Tau = \theta \cap \Tau$.
\label{def:cut}
\end{defn}

Informally, Def.~\ref{def:cut} ensures that a disc $\theta$ cuts the tunnel $\Tau$ in one place only, and it cuts it completely.

Having the discs cutting the tunnel defined, we can define how to generate cuts from the whole tunnel. Let $\Theta = \{\theta_i\}_{i=0}^{n}$ be a sequence of discs cutting tunnel $\Tau$. We require the cuts do not intersect in more than a single point, formally:
\begin{equation}
  x, y \in \Theta \Rightarrow |x \cap y| \leq 1
  \label{eq:intersect}
\end{equation}
Moreover, we need to upper-bound the distance between discs, so we can also upper-bound the movement of the ligand atoms (to allow a contiguous trajectory generation). Let $\delta$ be an upper-bound of discs' distance and $\theta_i, \theta_{i+1} \in \Theta$. Then formally we require:
\begin{equation}
  \begin{split}
    \forall x \in \theta_i \Rightarrow \exists y \in \theta_{i+1} \Rightarrow |x-y| \leq \delta \\
    \forall y \in \theta_{i+1} \Rightarrow \exists x \in \theta_{i} \Rightarrow |x-y| \leq \delta
  \end{split}
  \label{eq:upper_bound}
\end{equation}

The fundamental requirement is to move forward in the tunnel, \ie{} to generate a new cut ahead of the last cut:
\begin{equation}
  \theta_{i}, \theta_{i+1} \in \Theta \Rightarrow \langle \theta_i^{normal}, \theta_{i+1}^{center} - \theta_i^{center} \rangle > 0
  \label{eq:forward}
\end{equation}

Finally, we want to start at the first sphere and end at the last sphere:
\begin{equation}
  \qquad  \qquad S_1^{center} \in \theta_1
  \qquad  \qquad S_k^{center} \in \theta_n
  \label{cond:good_start}
\end{equation}

\subsection{Tunnel Discretization Computation}

The discretization algorithm iteratively adds new discs to $\Theta$. The tunnel geometry may be very complicated since the consecutive spheres may differ in radius significantly and may form sharp curves. Thus, we haven't found any simple analytical solution for the tunnel discretization. Instead, we have developed an iterative algorithm, which adds disc $\theta \in \Theta$ with the direction defined by a smoothed curve representing a tunnel and iteratively improve the positions of the disc to fulfill the requirements described in the previous section. In this section, we will focus on the main aspects of the algorithm, omitting the implementation details. 

\subsubsection{Direction in the Tunnel}
\begin{figure}[t]
\centering
\includegraphics[width=.4\hsize]{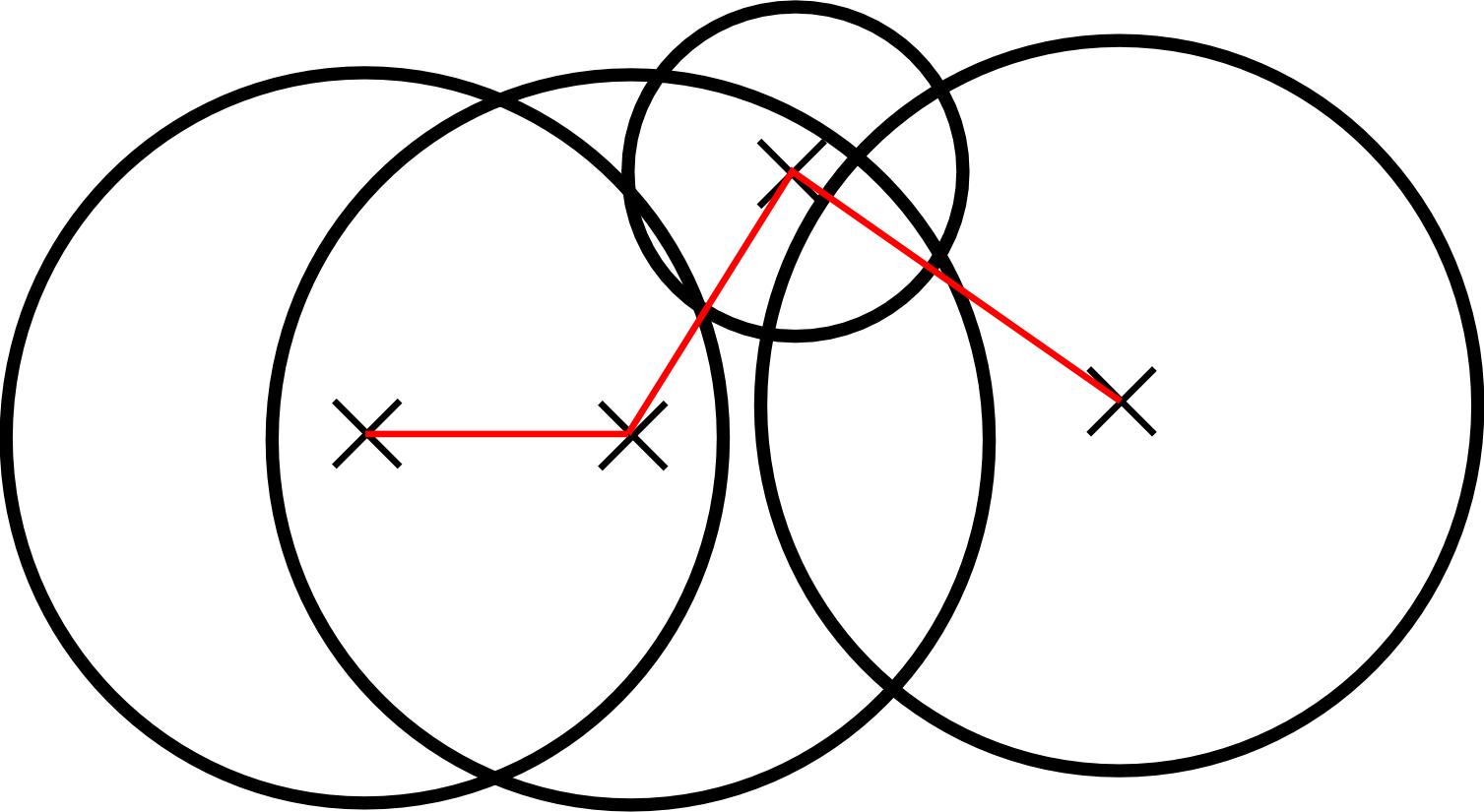}
\hspace{1cm}
\includegraphics[width=.4\hsize]{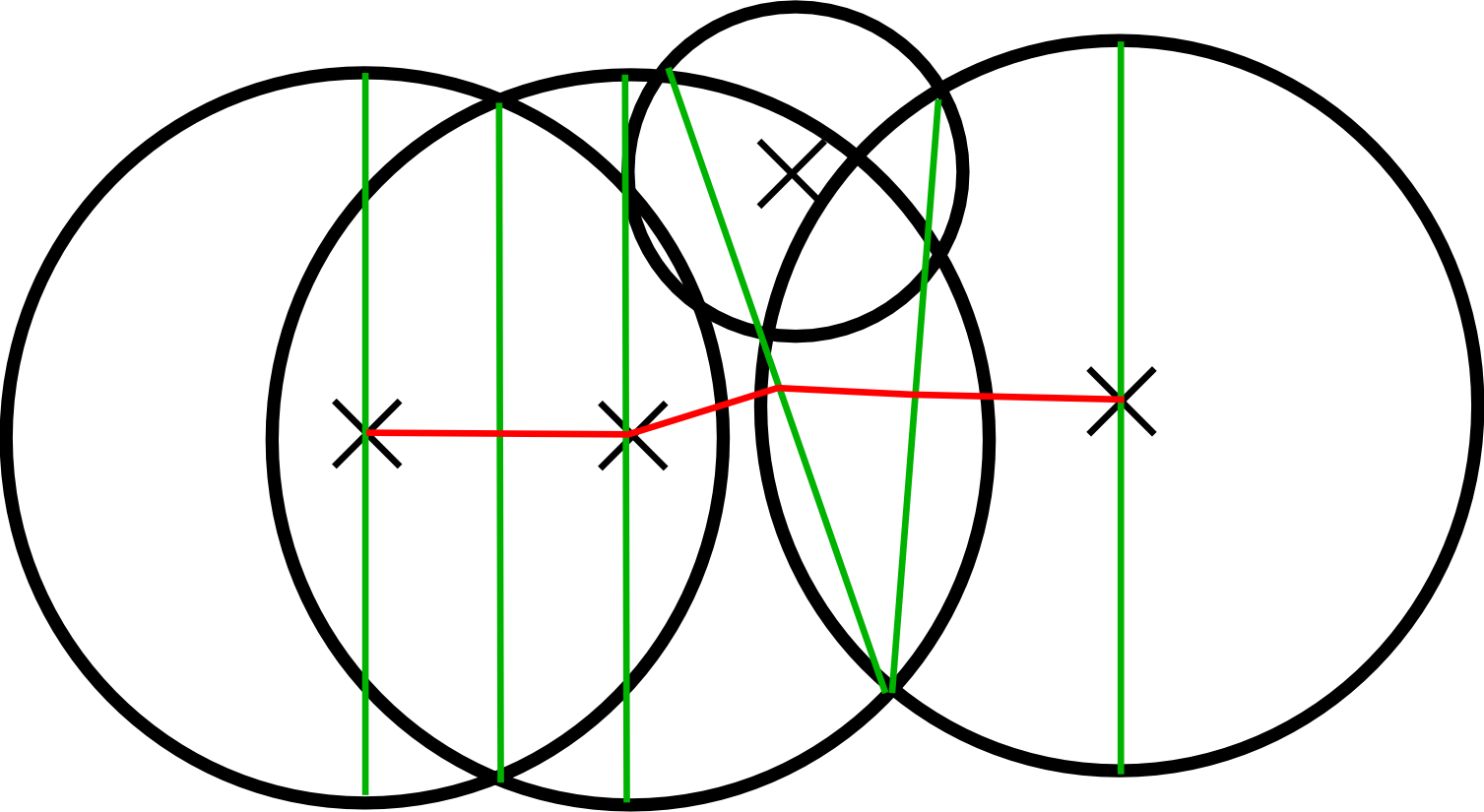}
\caption{Schematic 2D view of a curve (red) representing the tunnel direction. Left: naive solution where the centers of the spheres are connected; right: the centers of the minimal cuts (green) are connected.}
\label{fig:curve}
\end{figure}

First, we define a curve, which is used to guess the initial position of a newly placed disc. Such a curve should well-represent the direction of the tunnel. The easiest way is to connect the centers of spheres in $\Tau$. However, some spheres may have a small influence on the tunnel shape and create a curve which will not represent the tunnel direction well (see Figure~\ref{fig:curve} left). Thus, we compute the minimal cuts of the tunnel and connect the centers of those cuts (see Figure~\ref{fig:curve} right).

The curve defined by the centers of the cuts is not smooth, \ie{} its derivative does not exist in the cuts center. We further smooth the curve. The curve can be seen as a vector field and its smoothed variant as a weighted sum of the vector field, where the vector's weight quadratically descends with the growing distance between the points of the curve.

\subsubsection{Helper Functions}
Before we start with the tunnel discretization algorithm, we will introduce several important helper functions, which are used by the algorithm to place the discs along the curve representing the direction of the tunnel.

The function $\operatorname{fitDiscTunnel}$ computes the center and radius of the disc for the given plane $\rho$ defined by the normal $n$ and reference point $P$. The disc must be created to fulfill Definition~\ref{def:cut}: it must cut the tunnel at one place only and it must cut it completely. The function recursively builds a set of spheres $C \subset \Tau$, which contain $P$ or intersect both $\rho$ and some sphere in $C$. Having the $C$ constructed, the algorithm projects spheres from $C$ to $\rho$ and computes the circle encapsulating all projected spheres using the algorithm~\cite{fisher2005smallest}. The computed circle determines the center and the radius of the computed disc, the normal of the disc is the same as the normal of $\rho$.

The function $\operatorname{shiftDisc}$ modifies the disc $\theta_{i+1}$ to fulfill Conditions~\ref{eq:intersect},~\ref{eq:upper_bound} and~\ref{eq:forward} in relation to the already placed disc $\theta_{i}$. Let $\rho$ be a plane orthogonal to planes where $\theta_{i}$ and $\theta_{i+1}$ lies. In $\rho$, discs $\theta_{i}$, $\theta_{i+1}$ are projected as line segments. The algorithm is perfomed iterativelly: the line segment representing the disc $\theta_{i+1}$ is modified to not exceed $\delta$ in their ending points (Condition~\ref{eq:upper_bound}) and not intersect (Condition~\ref{eq:intersect}). After that, disc $\theta_{i+1}$ is reconstructed from the projection and the $\operatorname{fitDiscTunnel}$ is called (as Definition~\ref{def:cut} may be broken by shifting). This process is repeated untill there is no change on the disc $\theta_{i+1}$. Note that the $\operatorname{shiftDisc}$ always converges: in each iteration, the disc $\theta_{i+1}$ is moved closer to $\theta_{i}$. At some point, there is no violation of Conditions ~\ref{eq:intersect} and~\ref{eq:upper_bound}, and thus also no change is induced by $\operatorname{fitDiscTunnel}$.

In the case of a sharp curve in the tunnel, we need a more progressive placement of the disc, implemented in the function $\operatorname{shiftSharpTurn}$. The plane $\theta_{i+1}$ is initially placed in $\Delta$ distance from $\theta_i$, which breaks condition~\ref{eq:upper_bound}, but gives the idea of tunnel curvature. The function shifts the $\theta_{i+1}$ to intersect $\theta_{i}$ in the point nearest to $\theta_{i}$, sets its center and normal to not exceed $\delta$ distance from $\theta_{i}$ and calls $\operatorname{shiftDisc}$ to finalize the $\theta_{i+1}$ placement. The main difference between $\operatorname{shiftSharpTurn}$ and $\operatorname{shiftDisc}$ is that $\operatorname{shiftSharpTurn}$ sets the initial position of $\theta_{i+1}$ such that its normal is pointing to the further direction of the tunnel.

\subsubsection{Discretization Algorithm}

\begin{algorithm}
\caption{Algorithm for tunnel discretization}
\label{alg:digTunnel}
\begin{algorithmic}[1]

\Function{discretizeTunnel}{$\Tau, \delta$}
    \State $ centers \gets [ S^{center} \mid S \in \Tau ] $ \label{digTunnel:first_init}
    \
    \State $ discs \gets   [ \operatorname{fitDiscTunnel}(\operatorname{norm}(S_{1}^{center} - S_{0}^{center}), S_0^{center})] $
    \State $ curve \gets \operatorname{TunnelCurve}(centers) $ \label{digTunnel:last_init}
    \Statex

    \For{$ S_i \gets S_0, \dots, S_{|T| - 2} $} \label{digTunnel:main_loop}
        \State $ dir \gets S_{i + 1}^{center} - S_{i}^{center} $
        \State $ line \gets \{ S_{i}^{center} + t\cdot dir \mid t \in \mathbb{R} \} $

        \State $ d \gets 0 $
        \While {True} \label{digTunnel:inner_loop}
            \State $ prev\_disc \gets discs[|discs|-1] $
            \State $ plane \gets \operatorname{getPlane}(prev\_disc) $ \Comment{Construction of a plane containing disc}
            \State $ d \gets \dis(plane \cap line, S_{i}^{center}) + \epsilon  $
            \If {$ d > \norm{dir} $} \label{digTunnel:while_condition}
                \Break
            \EndIf
            \Statex

            \If{$\operatorname{makesSharpTurn}(prev\_disc, curve)$}
                \State $ disc\_center
                    \gets prev\_disc^{center} + \Delta * prev\_disc^{normal} $ \label{alg:shift_by_delta}
                \State $ disc\_normal \gets prev\_disc^{normal} $   \label{alg:same_normal}
                \State $ \operatorname{doShift} \gets \operatorname{shiftSharpTurn} $
            \Else
                \State $ disc\_center \gets prev\_disc^{center} + \epsilon * prev\_disc^{normal} $
                \State $ disc\_normal \gets \operatorname{getWeightedDir}(curve, i, d) $
                \State $ \operatorname{doShift} \gets \operatorname{shiftDisc} $
            \EndIf
            \State $ disc \gets \operatorname{fitDiscTunnel}(disc\_normal, disc\_center) $ \label{alg:fit_disk}
            \State $ disc \gets \operatorname{doShift}(prev\_disc, disc) $ \label{alg:shift_disk}

            \If{$ |discs| \geq 2  \wedge \dst(disc, discs[|discs|-2]) < \delta $} \label{digTunnel:last_if}
                \State $ \operatorname{Pop}(discs)  $ \label{digTunnel:last_if_end}
            \EndIf
            \State $ \operatorname{Append}(discs, disc) $
        \EndWhile
    \EndFor
    \State \Return $ discs $
\EndFunction
\end{algorithmic}
\end{algorithm}

The Algorithm~\ref{alg:digTunnel} presents the main structure of the tunnel discretization algorithm. {The input of the algorithm is a sequence of spheres $\Tau$ and the maximal distance between two discs $\delta$.} It uses helper functions, which are described in the previous section. The lines from~\ref{digTunnel:first_init} to~\ref{digTunnel:last_init} initialize the required data structures. The array $discs$ will be used to construct $\Theta$. During the initialization, the first disc is created with the same center as the first sphere and the normal given by the vector going from the center of the first to the center of the second sphere (so the first part of condition~\ref{cond:good_start} is fulfilled). The $curve$ contains the smoothed curve approximating the direction of the tunnel.

At line~\ref{digTunnel:main_loop}, the algorithm iterates over the spheres from $\Tau$, where in each step it constructs $line$, containing the line connecting the actual and the next sphere. In the inner loop, the discs are generated from the centre of the sphere $S_{i}$ to $S_{i+1}$ (condition at line~\ref{digTunnel:while_condition}). The variable $d$ determinates our distance from $S_{i}^{center}$ and controls the number of loop iterations at line~\ref{digTunnel:inner_loop}.

The function $\operatorname{isSharpTurn}$ detects if the tunnel forms a sharp turn at the particular place. If so, the algorithm uses a more aggressive strategy for the next disc placement: it places a new disc parallel to the current disc with distance $\Delta$. Otherwise, the center of the new disc is displaced by $\epsilon$, and its normal is set according to the weighted direction curve. The constant $\Delta$ determines the distance, which is checked for deciding whether the turn is sharp. We use $\Delta = 2 \delta$. The constant $\epsilon$ determines the granularity of discretization, and it must be lower than $\delta$. We set $ \epsilon = \frac{1}{10} \delta $.

\begin{figure}[t]
\centering
\includegraphics[width=.75\hsize]{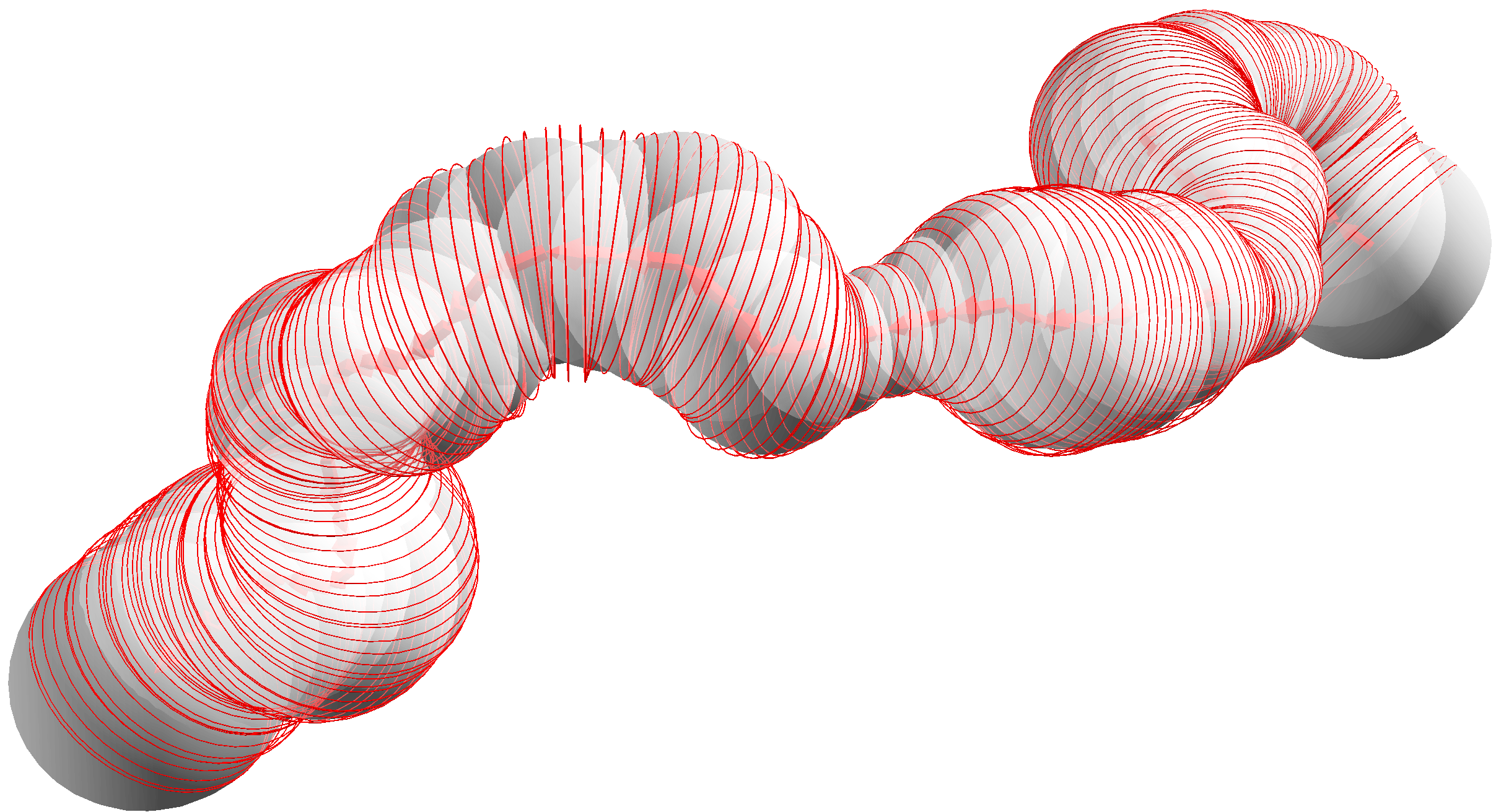}
\caption{Discretization of a tunnel in the native toluene/o-xylene monooxygenase hydroxylase. The red circles represent the discs, the red arrows represent the tunnel direction and the grey balls represent the tunnel obtained from Caver~\cite{chovancova2012caver}.}
\label{fig:tunnel_discretization}
\end{figure}

After the initial disc placement the function $\operatorname{fitDiscTunnel}$ is called. The function computes the center and radius of the disc, so the disc forms the cut of the tunnel (see Definition~\ref{def:cut}), and the radius of the disc is minimal. After fitting the disc, the function $\operatorname{doShift}$ may further improve its placement according to the curvature of the tunnel, so the disc will be placed to fulfill Conditions~\ref{eq:intersect},~\ref{eq:upper_bound} and~\ref{eq:forward}.
Finally, the algorithm checks if the disc created in the previous iteration can be omitted (to not generate too dense discretization) at lines from \ref{digTunnel:last_if} to \ref{digTunnel:last_if_end}. 
An example of the algorithm output is depicted in Figure~\ref{fig:tunnel_discretization}.

\section{Constrained Docking}
\label{sect:constrained_docking}

As we have described in Section~\ref{sect:oberview_constraint}, we are employing two types of constraints. Recall that the tunnel-position constraint snaps a selected atom $a_c$ in ligand $\lambda$ to the disc $\theta$ and the pattern constraint places the ligand $\lambda$ in the vicinity of $\lambda_{pattern}$. These constraints are implemented as new force-field terms added into the original AutoDock Vina force field. We first introduce the search-space optimization methods implemented in AutoDock Vina and after that describe how the newly-added constraints are implemented.

\subsection{AutoDock Vina Search Space Methods}

The molecular docking is an optimization problem, where the docking program is searching for the global minimum of energy defined by the position of the ligand and the flexible parts of the receptor with respect to the given force field.  Two optimization methods are working together in AutoDock Vina: a stochastic global optimization and a gradient-based local optimization.

The simulated system has multiple degrees of freedom (DoF), which must be searched. First, the ligand is considered as a body in the space, having its position and orientation vectors (six dimensions). Second, the ligand is a flexible body -- it may be bent by setting angular values for its free dihedral angles (one dimension per dihedral angle on every single bond). Third, the receptor may contain flexible side-chains (so also the receptor geometry may be partially flexible), where each flexible residue contains one or more free dihedral angles. Thus, the optimization algorithm must optimize a high number of DoF (typically tens). The nature of the chemical force field creates a lot of local minima in the search space.

The global-optimization method implemented in AutoDock Vina is based on the Markov chain Monte Carlo method (MCMC). The initial state (position and orientation of the ligand, dihedral angles of the ligand and flexible side chains) is selected randomly keeping the ligand within the defined box. Subsequently, a predefined number of global-optimization steps are performed. In a global optimization step, one or more DoF are changed by an upper-bound random value, so the newly-generated conformation of the ligand and flexible side-chains is in the upper-bound vicinity of the previous conformation. After the global optimization step, the local optimization is executed. If the local optimization converges to a better minimum than what was reachable from the previous global-optimization step, the global-optimization accepts the new step and uses it as a base for the next iteration. Otherwise, the new step is accepted only with small probability based on the Metropolis critorion (this feature allows the algorithm to escape from a local minimum). During the global-optimization, the significant local minima are stored, so AutoDock Vina is able to return multiple different conformations, not only the best one.

The local-optimization in AutoDock Vina implements the gradient-based Broy\-den-Fletcher-Goldfarb-Shanno (BFGS) method~\cite{wright1999numerical}. It is a variant of the Newton method, so derivatives of all force-field terms have to be computed (it also uses the second derivatives, but they are computed numerically).


\subsection{Tunnel-position Constraint}
To snap atom $a_c \in \lambda$ to the disc $\theta$, we add a force-field term which penalizes the ligand's positions where $|a_c - t| > 0$, where $t \in \theta: \forall u \in \theta, u \neq t, |a_c - u| > |a_c - t|$ (\ie{} the distance is computed as the distance between $a_c$ and the nearest point in $\theta$). The penalization energy $e_p$ is computed as a Gauss function of the distance $|a_c - t|$:

\begin{equation}
    e_{position} = p_{max} - p_{max}e^{-\frac{|a_c - t|^2}{0.5}}
    \label{eq:pos_constraint}
\end{equation}

where $p_{max}$ is the maximal value of the penalization energy. The constant $0.5$ used in the exponent has been selected experimentally. It ensures that the half of $e_{max}$ penalization is applied when $|a_c, t| = 0.5\,\angstrom$. Note that the bell-shaped function is used to avoid strong penalization of small distances between $a_c$ and $\theta$ too strongly in order to to keep the good numerical stability of the BFGS optimization method (so the Condition~\ref{eq:position} can be violated by a small distance in practice).

The term in Eq.~\ref{eq:pos_constraint} and its derivative has been added into the energy and force evaluation codes in AutoDock Vina, so it is applied to the dragged atom $a_c$ during the BFGS local optimization. Moreover, we have added a simple modification into the MCMC global optimization: when a new conformation is randomly generated, the ligand is shifted
by a vector $t-a_c$, so the global optimization step does not break the tunnel-position constraint. Note that the modification of the global optimization method is not necessary for applying the constraint in the docking, however, it speeds up the docking convergence.

\subsection{Pattern Constraint}
The pattern constraint keeps the ligand $\lambda$ in the vicinity of the pattern position $\lambda_{pattern}$, so it must be applied to all atoms of the ligand. The pattern constraint is also applied to the flexible side chains, however, for the sake of simplicity, we describe the application to a ligand only (the principle of the pattern is the same for the ligand and flexible side chains).

The pattern constraint is applied for all pairs of corresponding atoms $a \in \lambda$ and $b \in \lambda_{pattern}$. 
Let $\delta$ be the distance which is not penalized by the pattern constraint. The energy of the pattern constraint is computed as:

\begin{equation}
        e_{pattern} = c \cdot \sum_{a \in \lambda, b \in \lambda_{pattern}} max\left(0, |a-b| - \delta)\right)
        \label{eq:pat_constraint}
\end{equation}

where $c$ is a constant determining the strength of the pattern (it has been empirically set to 40). Apparently, Eq.~\ref{eq:pat_constraint} is not differentiable in the area where $|a-b| = \delta$. We define a derivative at these points to be $0$ and keep the computation of the pattern constraint simple for the sake of computational efficiency.

The MCMC global optimization method is constructed to perform a long chain of conformational changes to escape from the local minima. However, when the pattern constraint is applied, the movement of the ligand is restricted to the vicinity of the pattern. Thus, we have modified the global search, such that (i) the initial configuration mimics the position of the pattern and (ii) the number of steps of the global optimization is $100\times$ lower compared to the default setup. The MCMC method still allows to escape from the local minima but does not generate too long chains due to the limited range of ligand movements from the initial configuration.

Note that, as all constraints are evaluated as force field terms, they can be violated if there is some strong energy contribution generated by different force field term (\eg{} the pattern constraint may be violated when pushing the ligand against a rigid part of the receptor). Therefore, CaverDock filters the computed conformations and discard those that where the constraint violations exceeding some threshold (\eg{} if we consider contiguous conformation changes up to 0.5\,\angstrom, we can set the pattern constraint to penalize movement larger than 0.4\,\angstrom and tolerate the movement not exceeding 0.5\,\angstrom).

\section{Trajectory Search}
\label{sect:trajectory_search}

The implemented constraints allow us to define the ligand's position in the tunnel and upper-bound its distance from some pattern. Thus, it is possible to iteratively dock the ligand along the tunnel and analyze the energy of the transport process. Recall that we compute two types of trajectory:
\begin{itemize}
  \item lower-bound trajectory, which samples the tunnel finely, but the movement of the ligand and flexible side-chains is not contiguous;
  \item upper-bound trajectory, which is contiguous.
\end{itemize}
The lower-bound trajectory may underestimate the energy of barriers, as the ligand may flip or change its conformation dramatically between two consecutive steps (Figure~\ref{fig:lower-bound}). However, its computation is straightforward -- there is no dependence between consecutive steps (only the tunnel-position constraint is used), and therefore we may perform only $n$ docking steps, where $n$ is the number of discs. The upper-bound trajectory generates a contiguous movement of the ligand and side-chains using the pattern constraint. However, a high number of possible contiguous trajectories exist and there is no guarantee that our method finds the lowest energy trajectory. Thus, we call the trajectory upper-bound, as it is not known if its energy may be further improved. The optimal energies should lie between upper- and lower-bound values. 

Having a set of discs $\theta_1 \dots \theta_n$, the lower-bound trajectory is defined as 
\begin{equation}
\Lambda_{lb} = \lambda_{min}^1, \lambda_{min}^2, \dots, \lambda_{min}^n
\end{equation}
where $\lambda_{min}^i$ denotes the conformation at disc $i$ with the lowest energy from all explored $\lambda^i$. 

The upper-bound trajectory is defined as 
\begin{equation}
\Lambda_{ub} = \lambda_1^1,  \dots, \lambda_{m_1}^1, \lambda_1^2, \dots, \lambda_{m_2}^2, \dots, \lambda_1^n, \dots, \lambda_{m_n}^n
\end{equation}
where $m_1 \dots m_n \geq 1$, so the upper-bound trajectory follows a forward movement within the tunnel or changes the ligand position on a disc, but does not go backward.

\subsection{Ligand Movement Driving}

The trajectory search is driven by a set of final state automata. Each automaton is designed to perform different tasks:
\begin{itemize}
  \item \textit{general automaton}, controlling the overall progress of the trajectory search;
  \item \textit{lower-bound trajectory automaton}, performing lower-bound trajectory computation;
  \item \textit{forward movement automaton}, responsible for moving the ligand forward in the tunnel;
  \item \textit{optimization automaton}, optimizing the position of the ligand at the particular disc;
  \item \textit{backtracking automaton}, moving the ligand backward if it hits a barrier;
  \item \textit{convergence automaton}, pushing the ligand in a backtracked trajectory to converge with the forward trajectory
\end{itemize}
The execution of automata can be nested. For example, the general automaton calls the forward automaton to move forward in the tunnel, and the forward automaton calls the optimization automaton to improve the position on a disc. The reason for such an implementation is twofold. First, the different operations on the trajectory are separated and the code is easier to maintain. Second, the computation can be interrupted or altered at any place since each automaton performs at most one state transition per call. Thus, it is, for example, possible to execute multiple backtracking in parallel, as the general automaton can immediately continue after spanning a new backtracking automaton without waiting for the backtracking to finish.

The automata call all the constrained docking computations in a non-blocking manner. They submit tasks into an internal CaverDock queue, which is then processed in parallel. Therefore, it is possible to process multiple alternative trajectories in parallel by executing multiple automata in a simple serial loop, or an automaton may process multiple alternatives at once.

\subsubsection{Trajectory Search Parameters}
The algorithm for trajectory search uses several parameters, affecting its precision and a number of executed dockings (and hence the computation time). Those parameters, listed below, affect the state machines or the docking settings and may be configured by the user.
\begin{itemize}
  \item Parameter \textit{optimization stratregy} determines the optimization criterion for the trajectory search. In the current implementation, we can execute CaverDock to minimize the highest energy peak across the whole trajectory or minimize the integral of the trajectory energy.
  \item Parameter \textit{backtrack threshold} quantifies the energy difference (in kcal/mol) between the lower-bound and upper-bound energy which triggers the backtracking. It is expectable that the energy of contiguous upper-bound trajectory will be higher, however, too large difference may indicate that the upper-bound trajectory is suboptimal. Therefore, when the difference between the upper-bound and lower-bound trajectory energies at some disc exceeds the threshold, the backtracking is used to search for a better trajectory.
  \item Parameter \textit{backtrack limit} sets the number of discs that are processed before a new backtracking can be executed. The parameter may speed-up CaverDock when the number of executed backtrackings is too high. This parameter is ignored when the forward trajectory cannot be computed because of a bottleneck (the backtracking is then started immediately).
  \item Parameter \textit{contiguous threshold} sets the highest distance which the atoms can move between consecutive conformations, if this movement is considered contiguous.
  \item Parameter \textit{pattern limit} sets the highest distance that is not penalized by the pattern constraint. The pattern limit must be lower than the contiguous threshold so the pattern constraint may actually apply some force to ligand position before the ligand position is discarted.
\end{itemize}

\subsubsection{General Automaton}

\begin{figure}[t]
\centering
\includegraphics[width=.5\hsize]{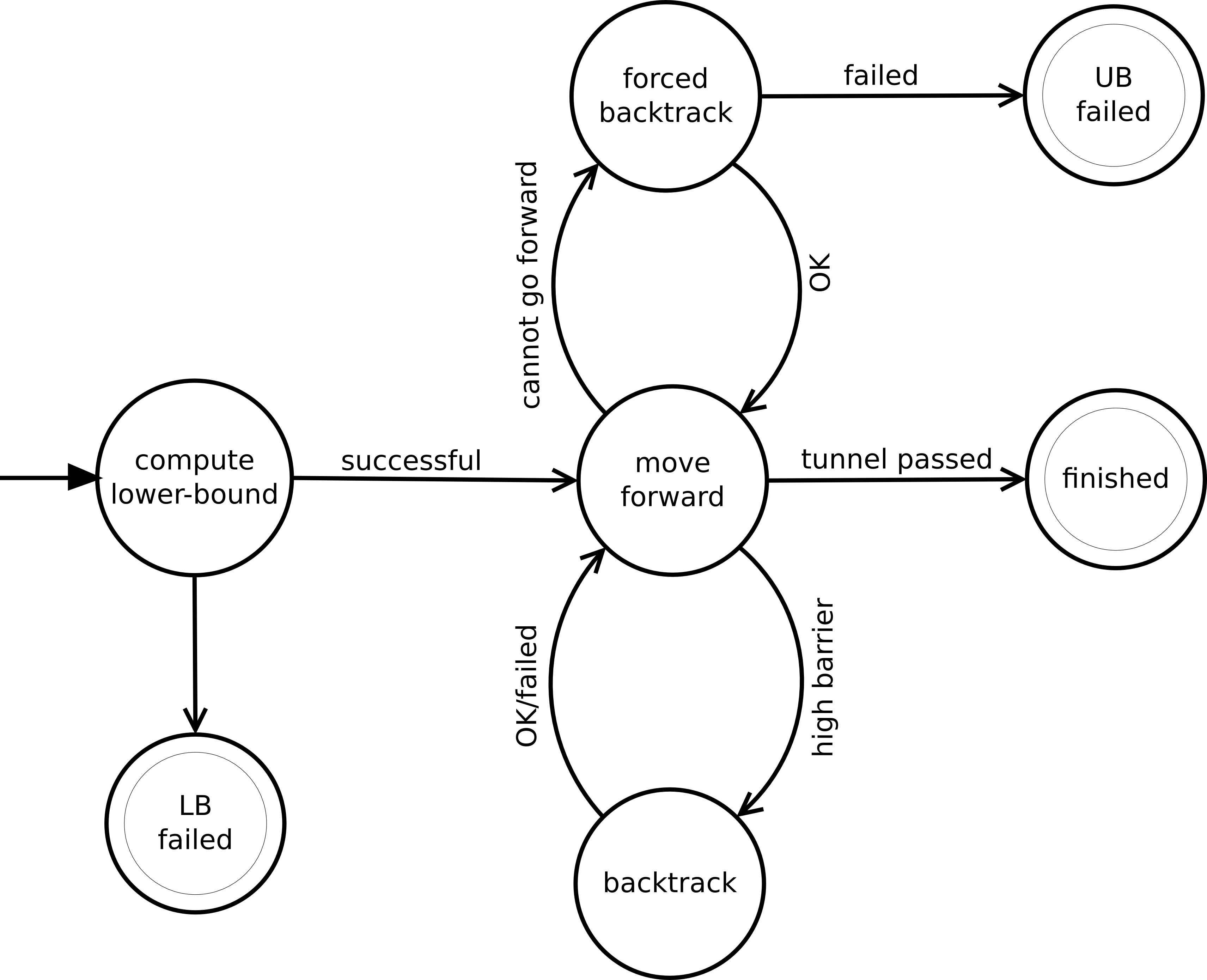}
\caption{General automaton.}
\label{fig:automaton_general}
\end{figure}

The simplified scheme of the general automaton is shown in Figure~\ref{fig:automaton_general}. After initialization, it starts to compute a lower-bound trajectory $\Lambda_{lb}$ and builds a cache of alternative conformations $\Lambda_{cache}$ (all examined conformations on discs $1 \dots n$). When the lower-bound computation is not successful (\ie{} $\exists i \in <1, n>, \lambda^i \notin \Lambda_{lb}$), the automaton halts in a \textit{LB failed} state. It may happen when the tunnel is very narrow in some part, and it is not possible to dock the ligand there. Otherwise, the general automaton starts with searching for a contiguous upper-bound trajectory. It inserts $\lambda_{min}^1$ into $\Lambda_{ub}$ and moves it into the \textit{forward} state performing the following steps:
\begin{itemize}
  \item Call the forward automaton to move forward from the last position $\lambda^i \in \Lambda_{ub}$ till it reaches the end of the tunnel, or requests backtracking.
  \item If the forward automaton requests backtracking (the energy of the forward trajectory is too high comparing to the lower-bound), create a backtracking automaton and change the state to \textit{backtrack}. The backtracking automaton starts from conformation $\lambda_j^{i+1} \in \Lambda_{cache}$, where $i$ is the last position in the forward trajectory $\Lambda_{lb}$ and $j$ is selected such that $\lambda_j^{i+1}$ has not been used for backtracking so far and its energy is minimal. It builds a backtrack trajectory $\Lambda_{backtrack}$. If the trajectory $\Lambda_{backtrack}$ is successfully found and improves the energy of the trajectory, it is implemented into $\Lambda_{ub}$. More precisely, the conformations in $\Lambda_{ub}$, from the conformation where the backtracking trajectory can be connected to the end of the trajectory, are removed and then $\Lambda_{ub} \leftarrow \Lambda_{ub} \cup \Lambda_{backtrack}$. Otherwise, a new backtracking is executed using a different starting conformation from $\Lambda_{cache}$, which has not been used for backtracking so far. If no such a conformation exists, then the general automaton returns to the \textit{forward} state.
  \item If the forward automaton requests a forced backtracking (it cannot find a forward trajectory), the backtracking automaton is created as in the previous case, and the general automaton changes its state to \textit{forced backtrack}. The difference to the \textit{forced backtrack} state is that $\Lambda_{backtrack}$ is implemented into $\Lambda_{ub}$ every time when it is found. If $\Lambda_{backtrack}$ cannot be found, the general automaton ends in \textit{UB failed} state: the upper-bound trajectory cannot be computed.
\end{itemize}

\subsubsection{Lower-bound Trajectory Automaton}
The automaton for the lower-bound is trivial. As there is no dependence between conformations in the lower-bound trajectory, the automaton computes the conformations for all discs. If some conformations are not computed, the automaton resubmits their computation once again (the docking process is stochastic, thus, it may fail occasionally). If there are still some conformations missing, the automaton ends in \textit{failed} state, otherwise, it ends in \textit{finished} state.

\subsubsection{Forward Automaton}

\begin{figure}[t]
\centering
\includegraphics[width=.3\hsize]{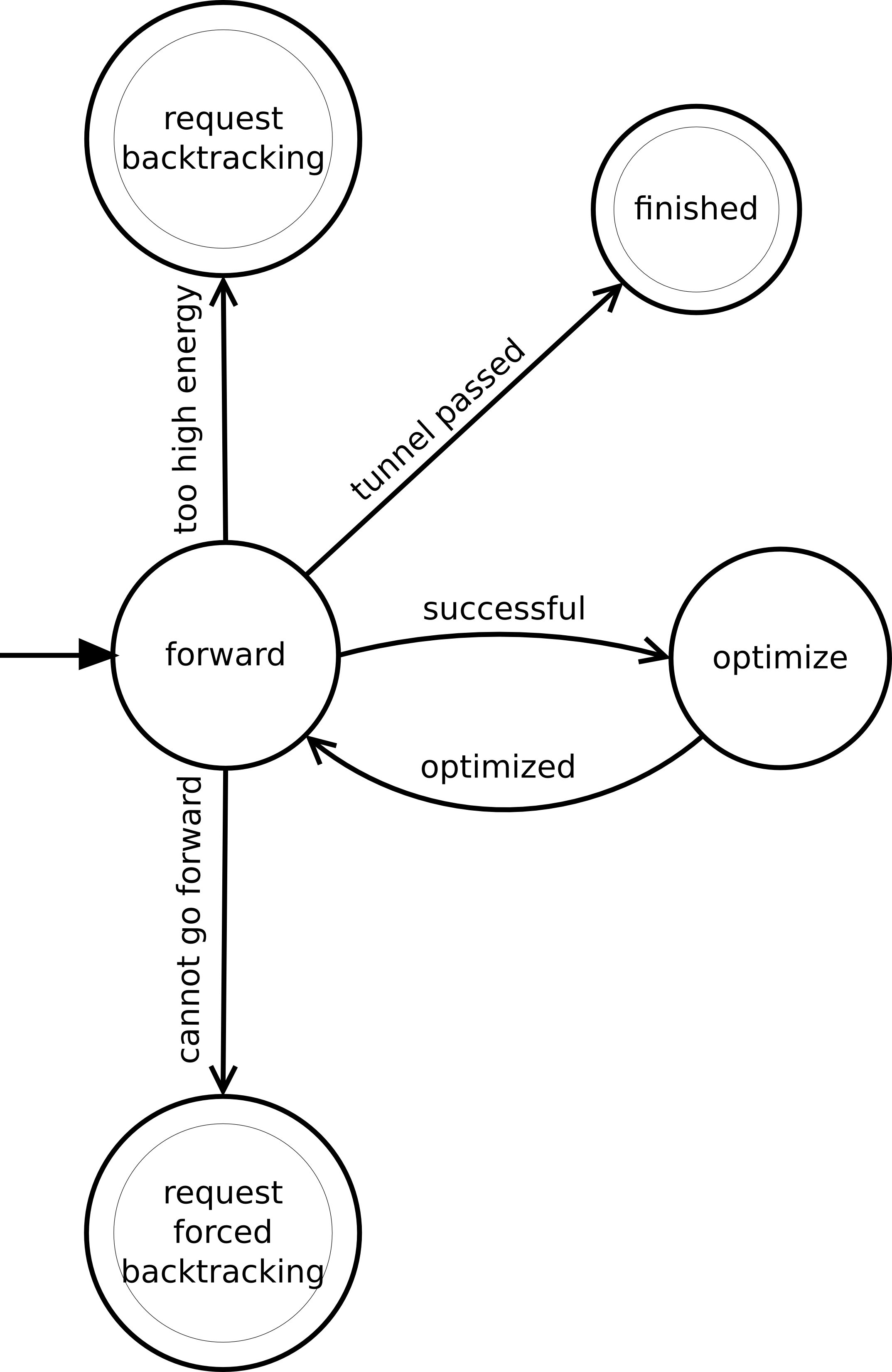}
\caption{Forward automaton.}
\label{fig:automaton_forward}
\end{figure}

The simplified forward automaton is depicted in Figure~\ref{fig:automaton_forward}. The automaton begins with the state $\lambda_{start}$ at disc $\theta_i$ and its purpose is to search for a contiguous sequence $\lambda_1^{i+1}, \dots, \lambda_{m_{i+1}}^{i+1}, \dots, \lambda_1^n, \dots, \lambda_{m_n}^n$. 

In the initial state \textit{forward}, the automaton searches for $\lambda_1^{i+1} \in \Delta \lambda_{start}^i$. If $\lambda_1^{i+1}$ is found and its energy is sufficiently low (does not differ from $\lambda_{min}^{i+1}$ by a value of \textit{backtrack threshold} parameter or more) or the backtracking was performed in recent steps (defined by \textit{backtrack limit}), the automaton moves to state \textit{optimize}. In the optimize state, the forward automaton executes an optimization automaton, which is responsible for searching a sequence $\Lambda_{opt} = \lambda_2^{i+1} \in \lambda_1^{i+1}, \lambda_3^{i+1} \in \lambda_2^{i+1}, \dots$, until the energy of the newly computed conformations is improved. After the optimization ends, the forward automaton returns to \textit{forward} state, sets $\Lambda_{ub} \leftarrow \Lambda_{ub} \cup \lambda_1^{i+1} \cup \Lambda_{opt}$ and sets the last conformation as the new starting conformation: $\lambda_{start} \leftarrow \lambda_{m_{i+1}}^{i+1}$, $i \leftarrow i+1$.

The conformation $\lambda_1^{i+1}$ may have too high energy. In such case, the forward automaton finishes in state \textit{request backtracking}, which indicates the general automaton that backtracking should be executed to improve the energy of the ligand trajectory. Moreover, the conformation $\lambda_1^{i+1}$ may not be found at all. In such case, the automaton ends in state \textit{request forced backtracking}, indicating the general automaton that the forward automaton cannot proceed further in the tunnel and the backtracking has to be used to find another ligand position.
If the automaton reaches the end of the tunnel, it finishes in state \textit{finished}, which indicates that a complete upper-bound trajectory has been found.

\subsubsection{Backtracking Automaton}

\begin{figure}[t]
\centering
\includegraphics[width=.5\hsize]{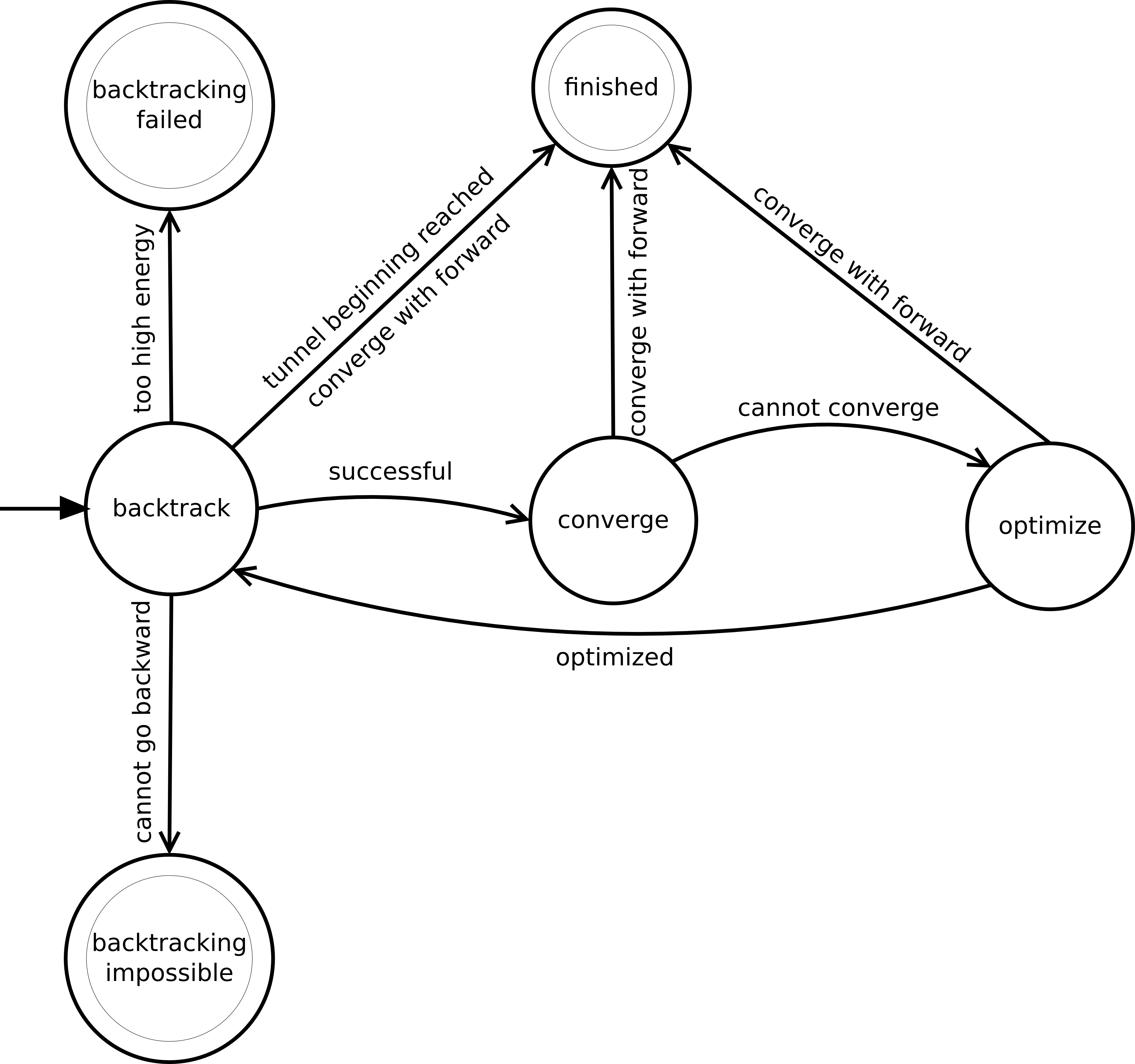}
\caption{Backtracking automaton.}
\label{fig:automaton_backtrack}
\end{figure}

The backtracking automaton is responsible for searching a trajectory from a selected backtracking point $\lambda_{start}^i$ at $\theta_i$ to the tunnel beginning, or to a point where the forward and backtracking trajectories converge and can be connected. The last conformation in the backtrack trajectory $\lambda_{last}^j$ can be connected to a forward trajectory, \ie{} $\lambda_{last}^j \in \Lambda_{backtrack}: \exists \lambda^{j-1} \in \Lambda_{ub}, \lambda_{last}^j \in \Delta \lambda^{j-1}$. 

The scheme of backtracking automaton is given in Figure~\ref{fig:automaton_backtrack}. The backtracking automaton begins in state \textit{backtrack}. In such state, it searches for $\lambda_1^{i-1} \in \Delta \lambda_{start}^i$. The following situations may occur:
\begin{itemize}
  \item If the new conformation cannot be found, the backtracking automaton ends in the state \textit{backtracking impossible}.
  \item If the backtracking is called to minimize the energy peak of the forward trajectory (defined by the parameter \textit{optimization strategy}), the energy of $\lambda_1^{i-1}$ is checked and if it leads to a higher energetic peak comparing to the corresponding part of the forward trajectory $\Lambda_{ub}$, the backtracking ends in the state \textit{backtracking failed}.
  \item Otherwise, $\lambda_1^{i-1}$ is inserted into $\Lambda_{backtrack}$ and the convergence state automaton is executed. The convergence automaton tries to find a trajectory connecting $\lambda_1^{i-1}$ to any conformation at the same disc in $\Lambda_{ub}$. If it is successful, the convergence trajectory is inserted into $\Lambda_{backtrack}$ and the state is changed to \textit{finished}. If convergence fails, the optimization automaton is executed analogically to the forward automaton and its trajectory is inserted into $\Lambda_{backtrack}$. If there is no convergence with $\Lambda_{ub}$ found during the optimization, the backtracking automaton returns back to \textit{backtrack} state and $\lambda^{start} \leftarrow \lambda^{i-1}_{last}$, where $\lambda^{i-1}_{last}$ is the last conformation from the optimization trajectory (at disc $\theta_{i-1}$) and $i \leftarrow i-1$.
\end{itemize}

In any of the \textit{backtrack}, \textit{convergence} and \textit{optimization} states, the backtracking and forward trajectories may converge (\eg{} the optimization step may push a ligand into the vicinity of any conformation from $\Lambda_{ub}$). Thus, the backtracking machine is checking each new state added to $\Lambda_{backtrack}$ and if it can be connected with the forward trajectory, it moves to a \textit{finished} state.



\subsubsection{Optimization Automaton}

The optimization automaton serves to improve the position at the same disc, building an optimization trajectory $\Lambda_{opt}$. It has two states only: \textit{optimization} and \textit{finished}. The automaton is executed with an initial position $\lambda^i_{start}$. In \textit{optimization} state, the new conformation $\lambda^i_{start+1} \in \Delta \lambda^i_{start}$ is searched. If the energy of $\lambda^i_{start+1}$ is better than that of $\lambda^i_{start}$, then it is inserted into $\Lambda_{opt}$, the automaton sets $\lambda^i_{start} \leftarrow \lambda^i_{start+1}$ and it stays in state \textit{optimization}, otherwise, it ends in state \textit{finished}.

\subsubsection{Convergence Automaton}

The convergence automaton optimizes the ligand position at the same disc similarly to the optimization automaton. However, instead of moving the ligand to the local minimum, it searches for a trajectory beginning at $\lambda^i_{start}$ to $\lambda^i_{dest}$. It uses a soft pattern constraint to attract the ligand atoms to a position determined by $\lambda^i_{dest}$ (by setting the constant $c$ in Equation~\ref{eq:pat_constraint} to one, instead of 40 used in the constraint forcing the contiguous movement). 

The convergence automaton has three states: \textit{convergence}, \textit{finished} and \textit{failed}. In the \textit{convergence} state, the automaton searches for $\lambda^i_{start+1} \in \Delta \lambda^i_{start}$, where the soft pattern constrain attracts $\lambda^i_{start+1}$ to $\lambda^i_{dest}$. If the average distance of atom pairs between $\lambda^i_{start+1}$ and $\lambda^i_{dest}$ is the same or bigger comparing to the distance between $\lambda^i_{start}$ and $\lambda^i_{dest}$, the automaton ends in state \textit{failed}. It ends in the \textit{failed} state when the energy added by the convergence trajectory would result in the end of the backtracking process. Otherwise, it checks if $\lambda^i_{start+1} \in \Delta \lambda^i_{dest}$: if the condition holds, it ends in state \textit{finished}, otherwise, it continues in state \textit{convergence} by setting $\lambda^i_{start} \leftarrow \lambda^i_{start+1}$.

\subsection{Software Architecture}

The CaverDock is built as an MPI application using master-slave parallelism. There is one master process, driving the trajectory search (\ie{} executing automatons and assigning work for slaves). The slave processes are responsible for computing the constrained docking: they receive constraints from the master (position of the disc and position of pattern atoms) and send the computed conformations with the computed energies (chemical force field and constraints' energy). 

The master process runs in a loop, querying automatons and gathering data from slaves. The automatons are called in a non-blocking fashion: they submit a work package describing the input for the docking. This work-package is held by the master process and assigned to a slave when is ready. Therefore, automatons can submit any number of work packages without waiting for the result and react by changing its state when the work is completely done.

The CaverDock can also be executed in a simple docking mode. In such case, only one slave process is executed, taking the input for the docking from a command line. When the user executes CaverDock with the parameters used in the original AutoDock Vina, CaverDock operates exactly as AutoDock Vina. However, it is possible to also pass the constraint via command line. Therefore, CaverDock is usable also as a docking tool allowing richer control over the docking process via constraints. It may be applied for observing a particular docking conformation when the user is interested in searching for a ligand conformation with some atoms restricted in a defined area of their interest.

\section{Evaluation}
\label{sect:evaluation}

In this section we compare the results of CaverDock with similar tools and demonstrate CaverDock's ability to analyze complex tunnels in reasonable time on chemically-relevant data. We assume in this study that the AutoDock Vina force field used in CaverDock returns realistic energy values and therefore we do not compare CaverDock's results with experimental data. The evaluation proving the chemical relevance of the computed results is being presented in parallel in other papers~\cite{vavra2018caverdock, pinto2018fast}.

\subsection{Comparison with Similar Tools}

\begin{figure}[t]
\centering
\includegraphics[width=.6\hsize]{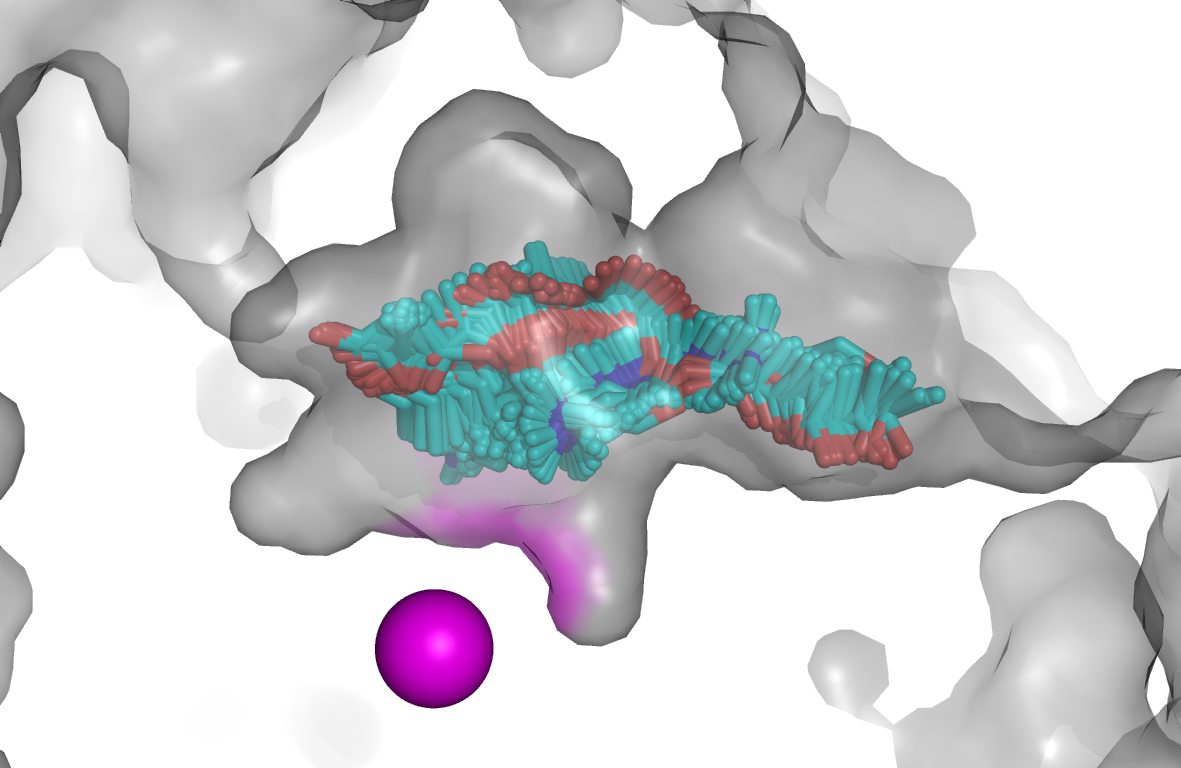}
\hspace{1cm}
\includegraphics[width=.2\hsize]{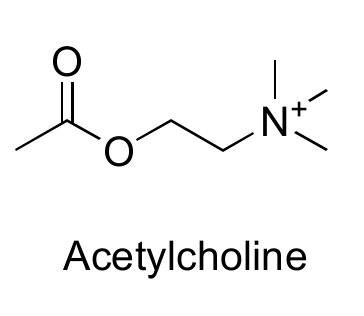}
\caption{Trajectory of acetylcholine in the tunnel of acetylcholinesterase computed by CaverDock. All 174 positions of acetylcholine are shown as the superimposed cyan sticks. The surface of the protein (close-up of tunnel) is shown as the grey surface, the catalytic S203 ($C\alpha$ atom) as the magenta sphere, and the chemical structure of this ligand is shown on the right.}
\label{fig:cd_traj}
\end{figure}

The comparison with SLITHER~\cite{lee2009slither} and MoMA-LigPath~\cite{devaurs2013moma} tools is given in this section. We demonstrate the qualitative difference in the produced trajectories using an example of the transportation of acetylcholine through a tunnel in the protein acetylcholinesterase (PDB ID 1MAH). The acetylcholine has been moved through the tunnel from the active site to the protein surface. The trajectory computed by CaverDock is shown at Figure~\ref{fig:cd_traj}. It can be seen that there are no gaps (empty spaces) in acetylcholine trajectory -- the movement of its atoms is contiguous.

\begin{figure}[t]
\centering
\includegraphics[width=.49\hsize]{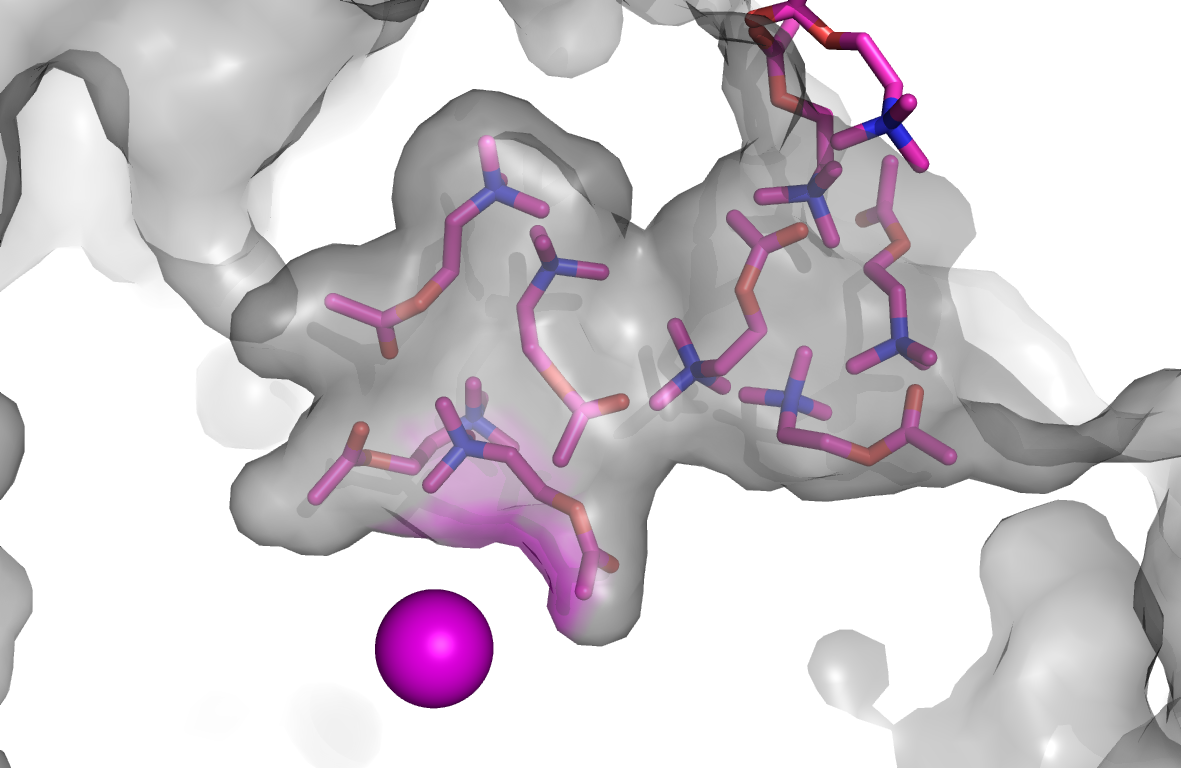}
\includegraphics[width=.49\hsize]{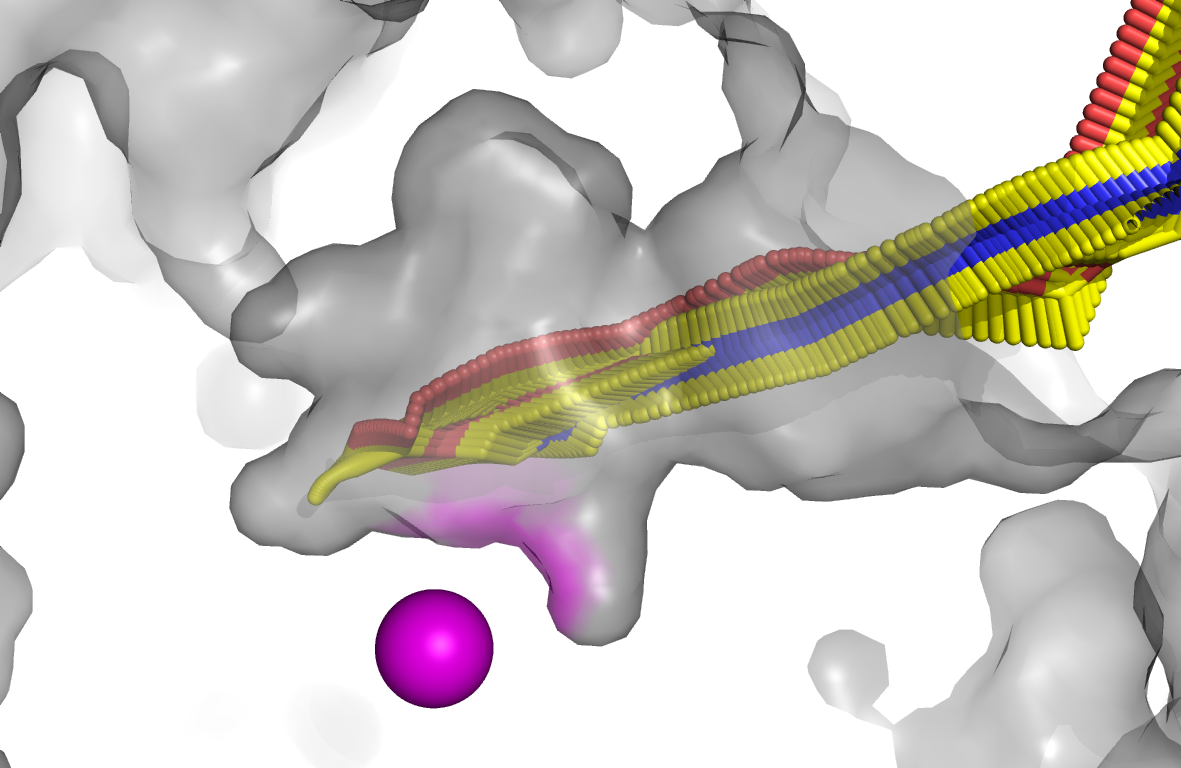}
\caption{Trajectory of acetylcholine in the tunnel of the acetylcholinesterase computed by SLITHER (left image) and MoMA-LigPath (right image). The surface of the protein (close-up of tunnel) is shown as the grey surface, the catalytic S203 (C atom) as the magenta sphere.}
\label{fig:slmm_traj}
\end{figure}

The trajectories computed by SLITHER and MoMA-LigPath are shown in Figure~\ref{fig:slmm_traj}. SLITHER does not implement a constrained docking, and as a consequence it produces large gaps in the computed trajectory. MoMA-LigPath, on the other hand, produces a contiguous trajectory. However, no chemical force field is used in MoMA-LigPath, so the user has no information describing the energy profile associated with the trajectory. Moreover, the trajectory does not reflect the chemical interactions and therefore can follow a path which would not be favored in the real systems. It can also be seen that the trajectory produced by MoMA-LigPath is more regular compared to CaverDock (the direction of atoms' movements is similar in multiple following conformations), which is very likely a consequence of the missing chemical forces, which increase the complexity of the optimized trajectory but are essential to describe the realistic behavior of the molecules.

\subsection{CaverDock Runtime and Robustness}

For testing the stability and time demands of CaverDock, we have prepared a representative set of biologically relevant protein-ligand pairs shown in Table~\ref{tab:complexes}. The set contains proteins with both short and long tunnels (\eg{} the insulin hexamer tunnel is discretized to 42 discs only, whereas the glucose transporter tunnel is discretized to 362 discs). The complexity of ligands also heavily varies: phenol has only 7 DoF (6 for the position and orientation and 1 free dihedral angle) whereas the arachidonic acid has 20 DoF (14 free dihedral angles).

\begin{table}
    \caption{The experimental set of molecules used for CaverDock evaluation.}
    \label{tab:complexes}
        \centering
        \small
        \begin{tabular}{|l|l|l|l|}
        \hline
            protein             & ligand     & ligand DoF   & discs\\
            \hline
        haloalkane dehalogenase     & 1-chlorobutane    & 8        & 72 \\ 
        acetylcholinesterase    & acetylcholine    & 10        & 85 \\ 
        leucine transporter        & leucine    & 10        & 105 \\ 
            lactose permease        & lactose    & 18        & 138 \\ 
        glucose transporter     & glucose    & 12        & 362 \\
        lipase B             & 4-methyloctanoic acid    & 12    & 60 \\
        insulin hexamer         & phenol    & 7        & 42 \\ 
        aquaporin Z            & glycerol    & 11        & 121 \\ 
        vitamin D receptor        & 1,25-dihydroxyvitamin D3 & 13    & 92 \\ 
            cytochrome P450 2E1        & arachidonic acid & 20        & 149 \\ 
        \hline
       \end{tabular}
\end{table}

We have tested CaverDock runtime using desktop computer equipped by AMD Ryzen 7 1700 (8 cores at 3.0\,GHz) and 16\,GB RAM. The resulting time, quality of the result and a number of performed docking calculations are shown in Table~\ref{tab:timing}.

\begin{figure}[t]
\centering
\includegraphics[width=.5\hsize]{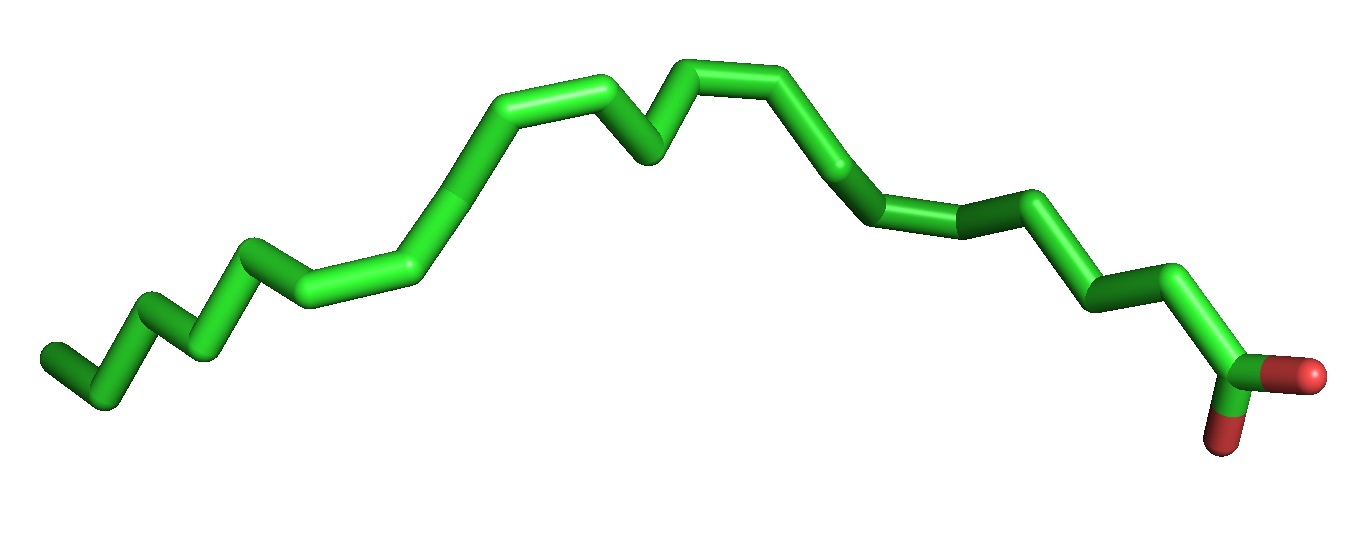}
\caption{Arachidonic acid.}
\label{fig:ligands}
\end{figure}

CaverDock was not able to compute the upper-bound trajectory in two cases (vitamin D receptor + 1,25-dihydroxyvitamin D3 and cytochrome P450 2E1 + arachidonic acid), whereas the lower-bound trajectory has been computed in all of the tested cases. The arachidonic acid contains a long chain with a high number of DoF (see Figure~\ref{fig:ligands}), which complicates the process of searching for the upper-bound trajectory. We suppose that CaverDock heuristics fails to find the contiguous movement of such a complicated molecule. 1,25-dihydroxyvitamin D3 has also a high number of DoF, but the main reason why CaverDock failed to compute contiguous trajectory is due to the narrow part of the tunnel entrance, which is difficult to pass with a contiguous movement. In contrast, lactose has also a high number of DoF, but the tunnel in the lactose permease is wider and CaverDock had no problem to compute the contiguous upper-bound trajectory.

The computation time ranges from 1\,m3\,s to nearly 2.5\,h. The CaverDock heuristic is in $\mathcal{O}(n^2)$, where $n$ is the number of discs (as backtracking can be issued during the whole trajectory and, in the worst case, may continue to the trajectory beginning). Therefore, the number of docking calculations, and hence the computational time, may grow quadratically with the tunnel length. However, the backtracking is not issued often in narrow tunnels, where the running time may grow according to the number of discs. The time required for each docking is highly dependent on the number of DoF, for example, 14.3 dockings per second are computed in the case of phenol (7 DoF), but only 1.33 dockings per second are computed in the case of lactose (18 DoF).

\begin{table}
    \caption{The output characteristics and computational demand of CaverDock calculations using ten different biological systems. LB = lower-bound, UB = upper-bound.}
    \label{tab:timing}
        \centering
        \small
        \begin{tabular}{|l|l|l|l|}
        \hline
            protein             & result     & runtime   & dockings\\
            \hline
            haloalkane dehalogenase & LB+UB     & 2m14s     & 1,824 \\ 
            acetylcholinesterase    & LB+UB     & 4m46s     & 1,920 \\ 
            leucine transporter     & LB+UB     & 7m4s      & 2,688 \\ 
            lactose permease        & LB+UB     & 67m20s    & 5,384 \\ 
            glucose transporter     & LB+UB     & 149m47s   & 23,888 \\
            lipase B                & LB+UB     & 3m19s     & 896 \\
            insulin hexamer         & LB+UB     & 1m3s     & 900 \\ 
            aquaporin Z             & LB+UB     & 8m50s     & 3,896 \\ 
            vitamin D receptor      & LB only   & 40m6s     & 4,132 \\ 
            cytochrome P450 2E1     & LB only   & 19m35s    & 644 \\ 
        \hline
       \end{tabular}
\end{table}


We have tested the set of molecules described in Table~\ref{tab:complexes} also with SLITHER and MoMA-LigPath. We were not able to compute the trajectories in the tunnels of  lipase B, insulin hexamer, aquaporin Z, vitamin D receptor and cytochrome P450 2E1 with SLITHER, and in glucose transporter, aquaporin Z, vitamin D receptor and cytochrome P450 2E1 with MoMA-LigPath. Therefore, at least for our testing set, CaverDock was more robust regarding its ability to compute the trajectories.  The runtime of SLITHER and MoMA-LigPath is in order of minutes in the worst case. Therefore, CaverDock time is comparable for simpler cases but may be significantly higher when a large number of dockings needs to be executed.

\subsection{Energy Profiles}

\begin{figure}[t]
\centering
\includegraphics[width=.32\hsize]{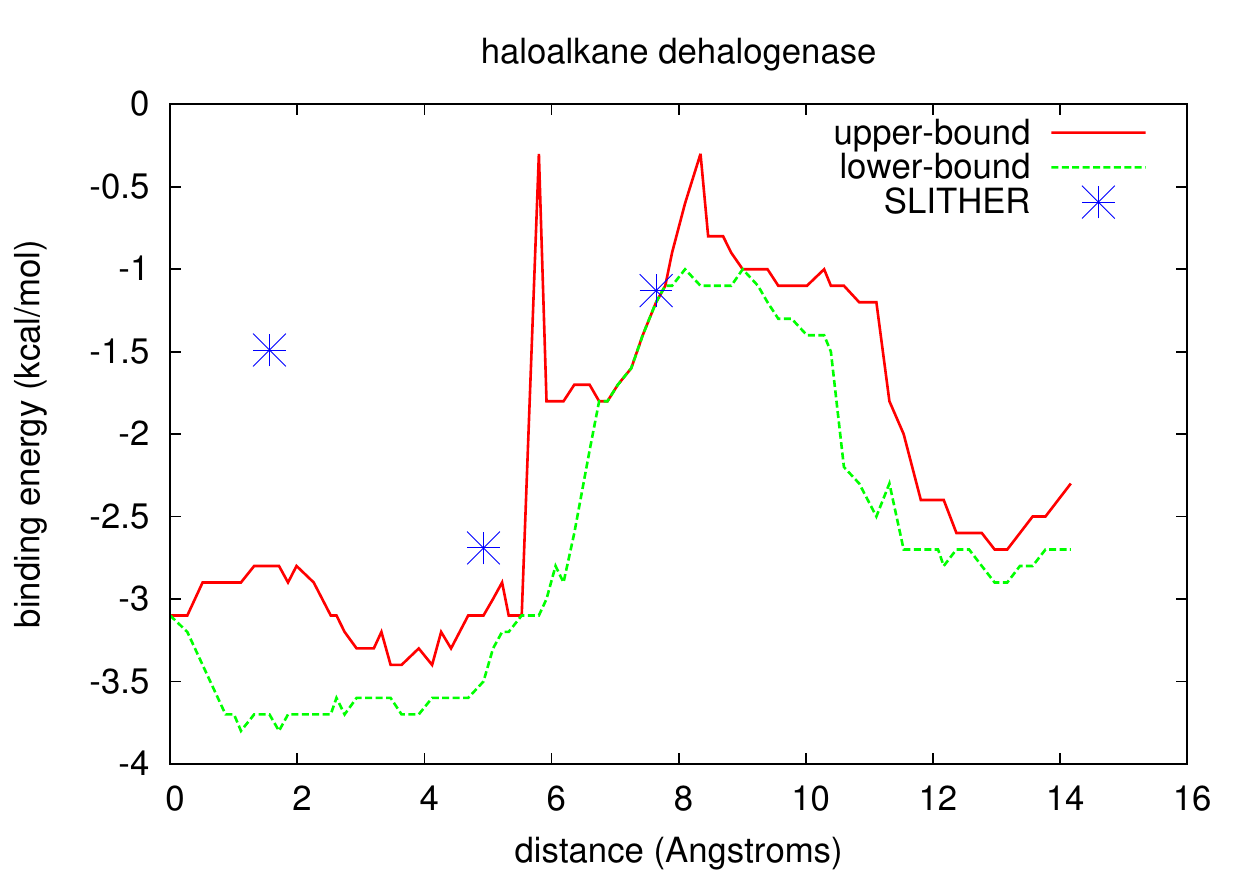}
\includegraphics[width=.32\hsize]{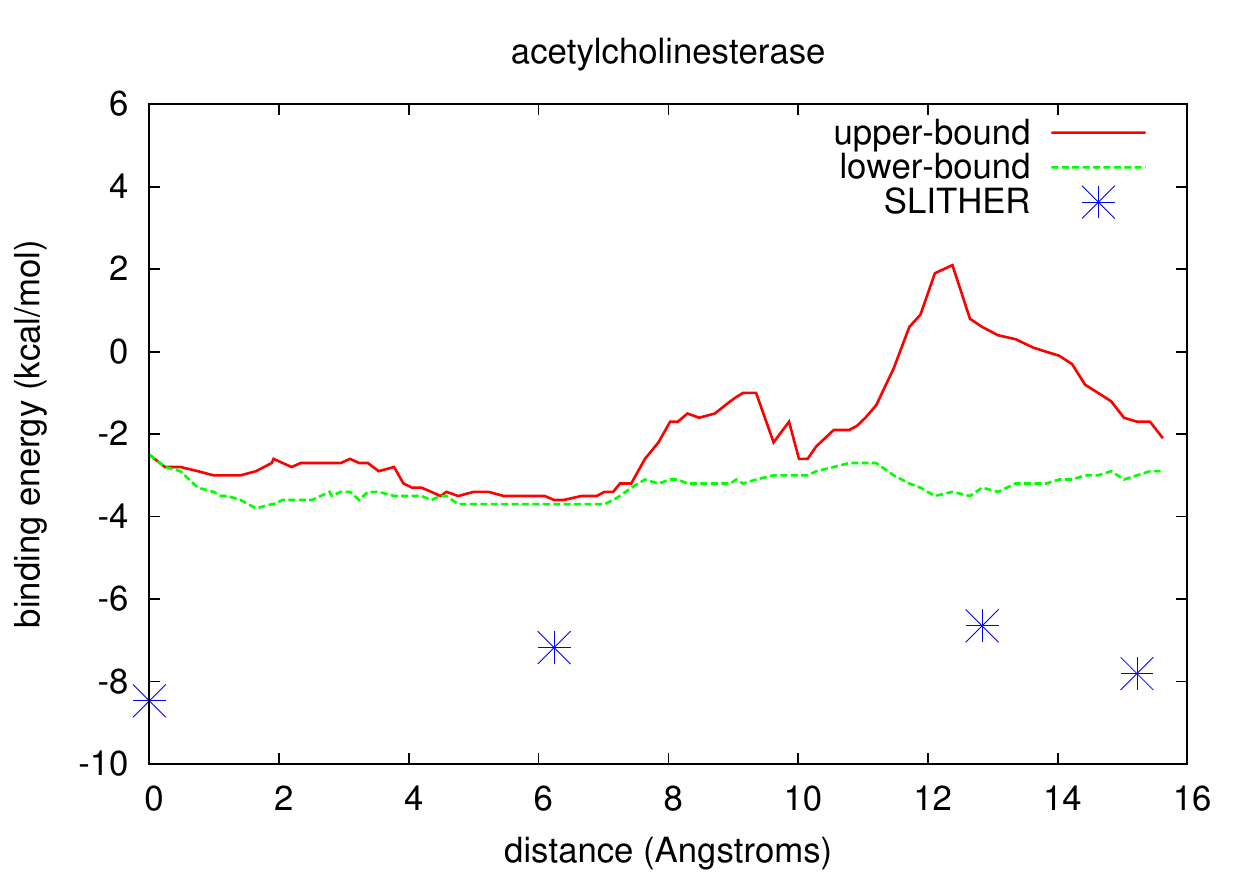}
\includegraphics[width=.32\hsize]{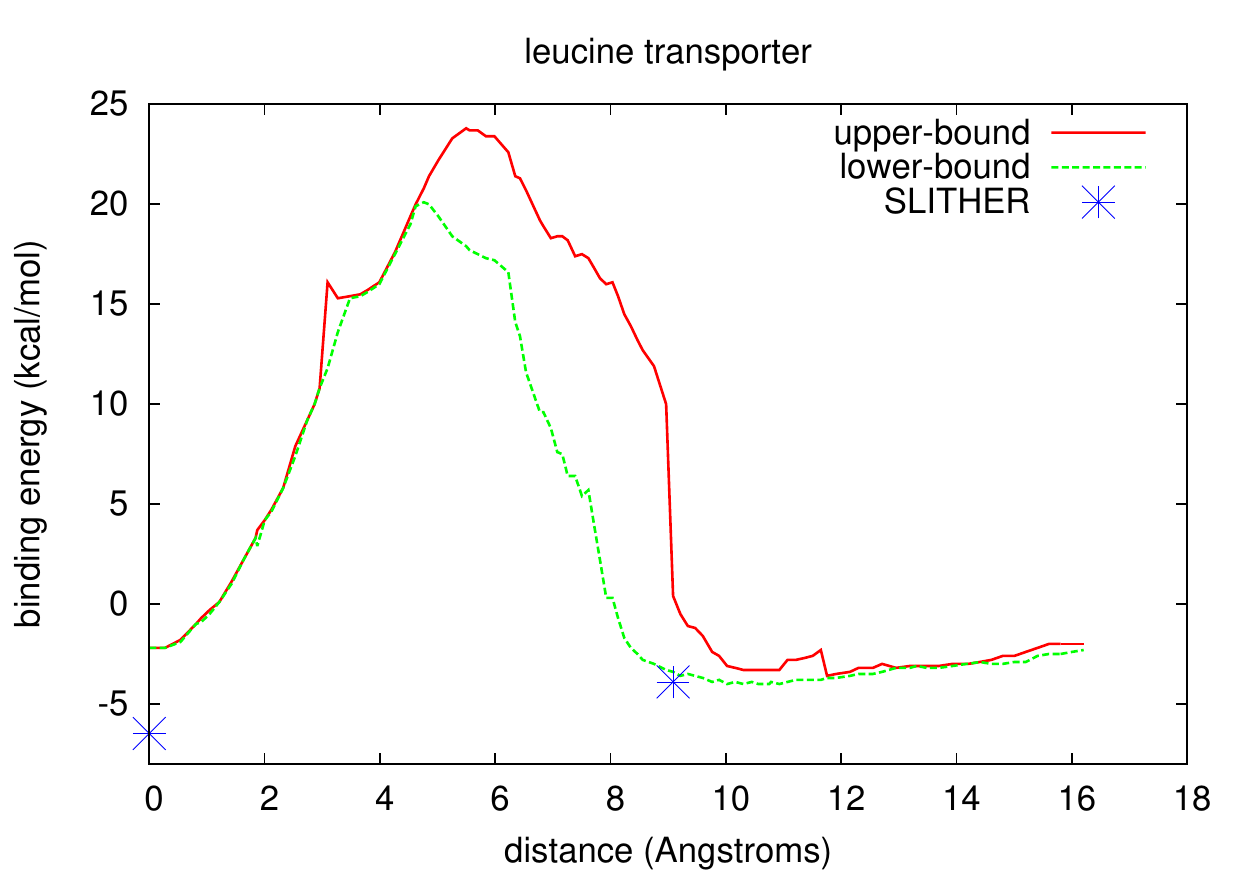}

\includegraphics[width=.32\hsize]{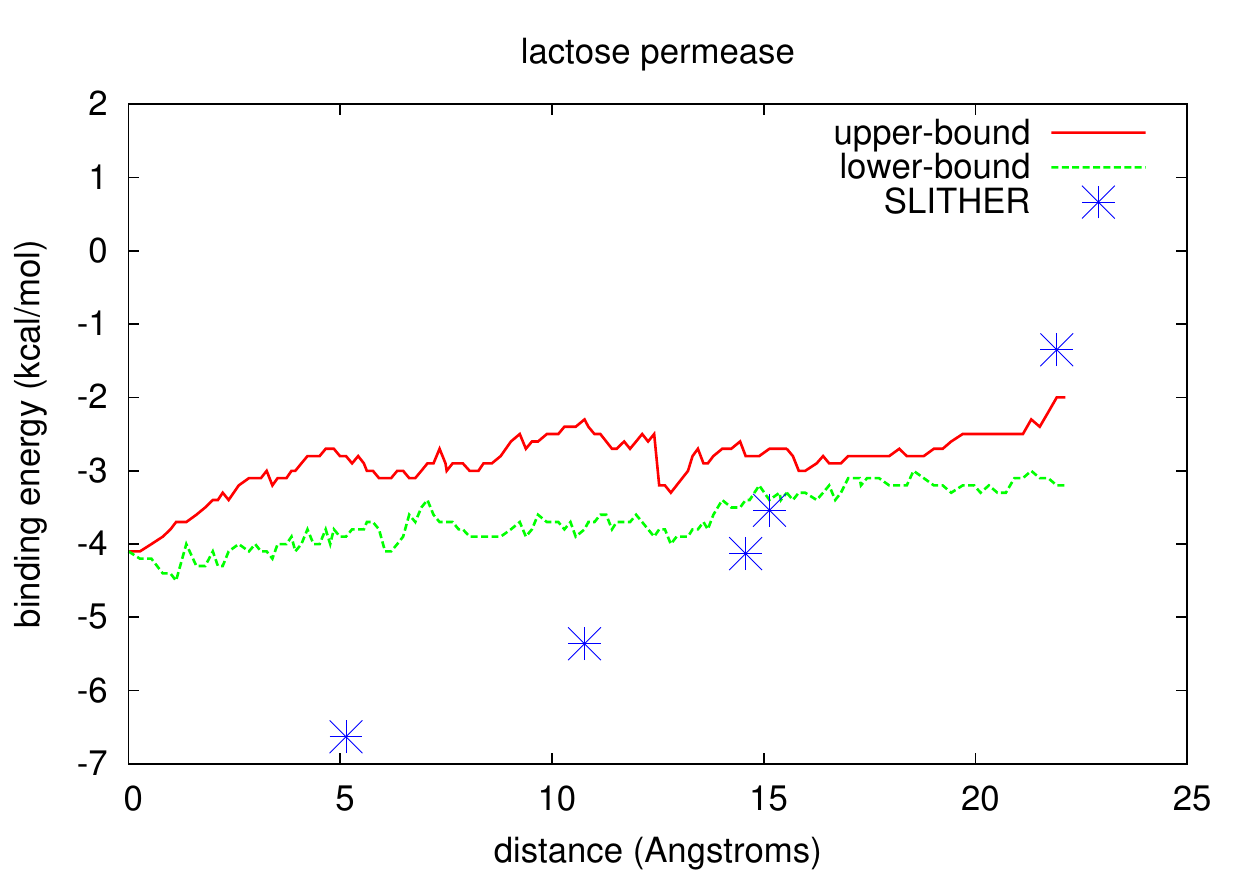}
\includegraphics[width=.32\hsize]{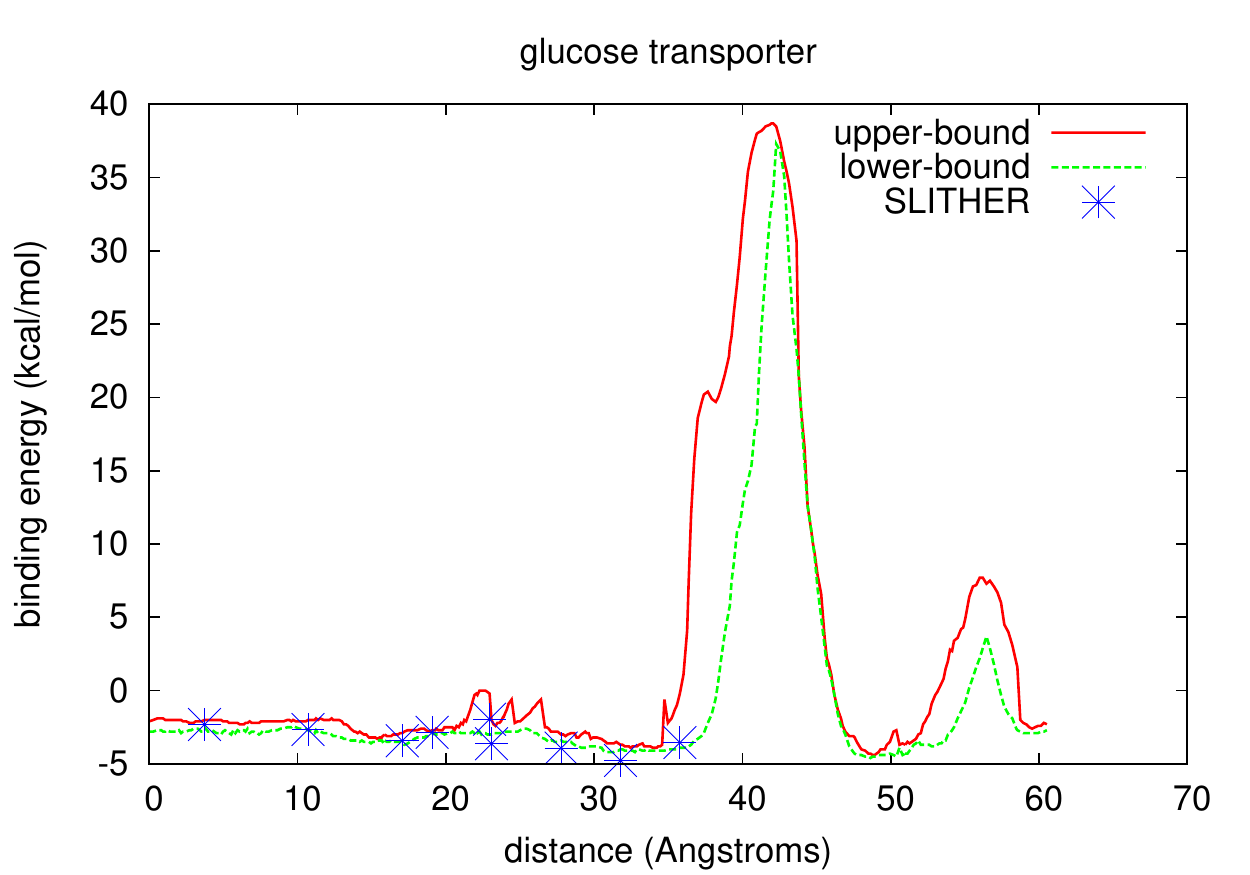}
\includegraphics[width=.32\hsize]{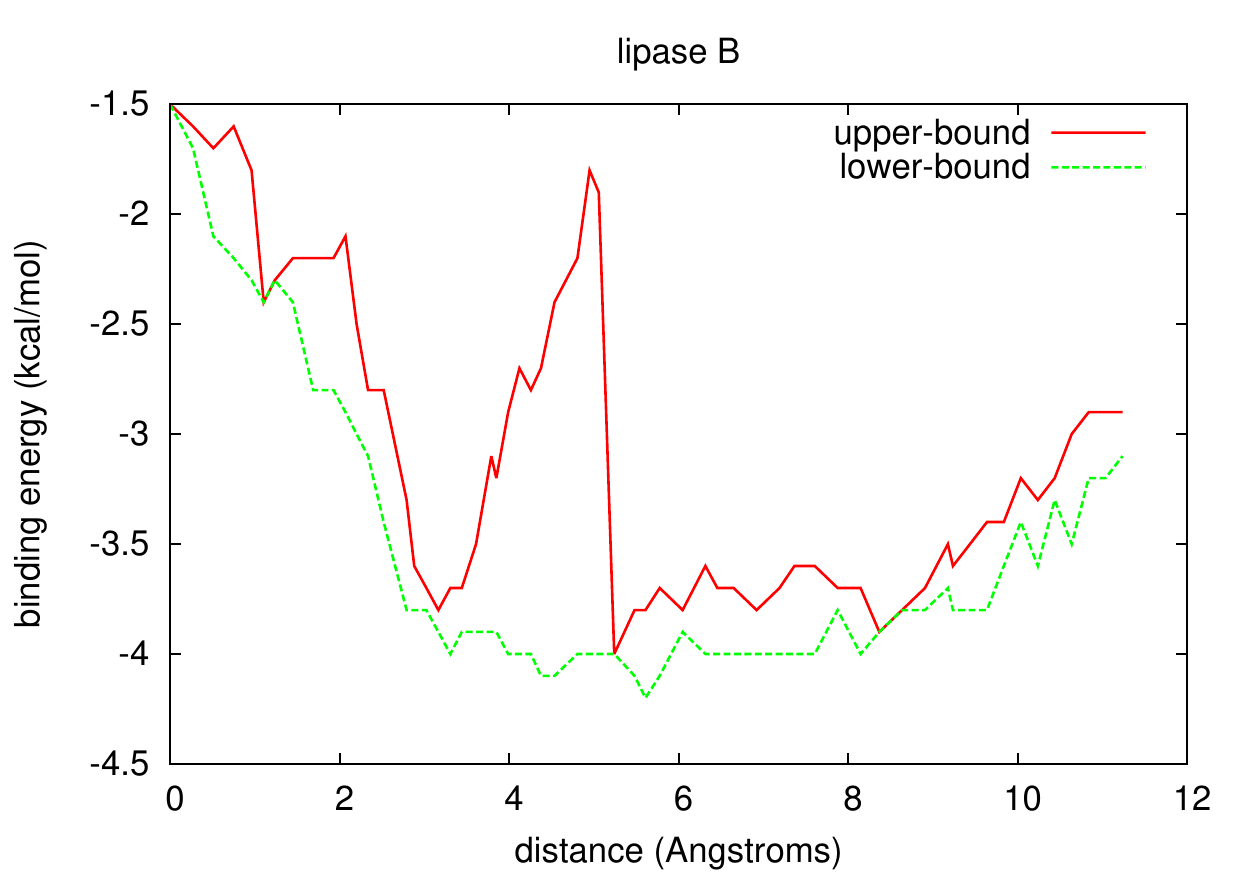}

\includegraphics[width=.32\hsize]{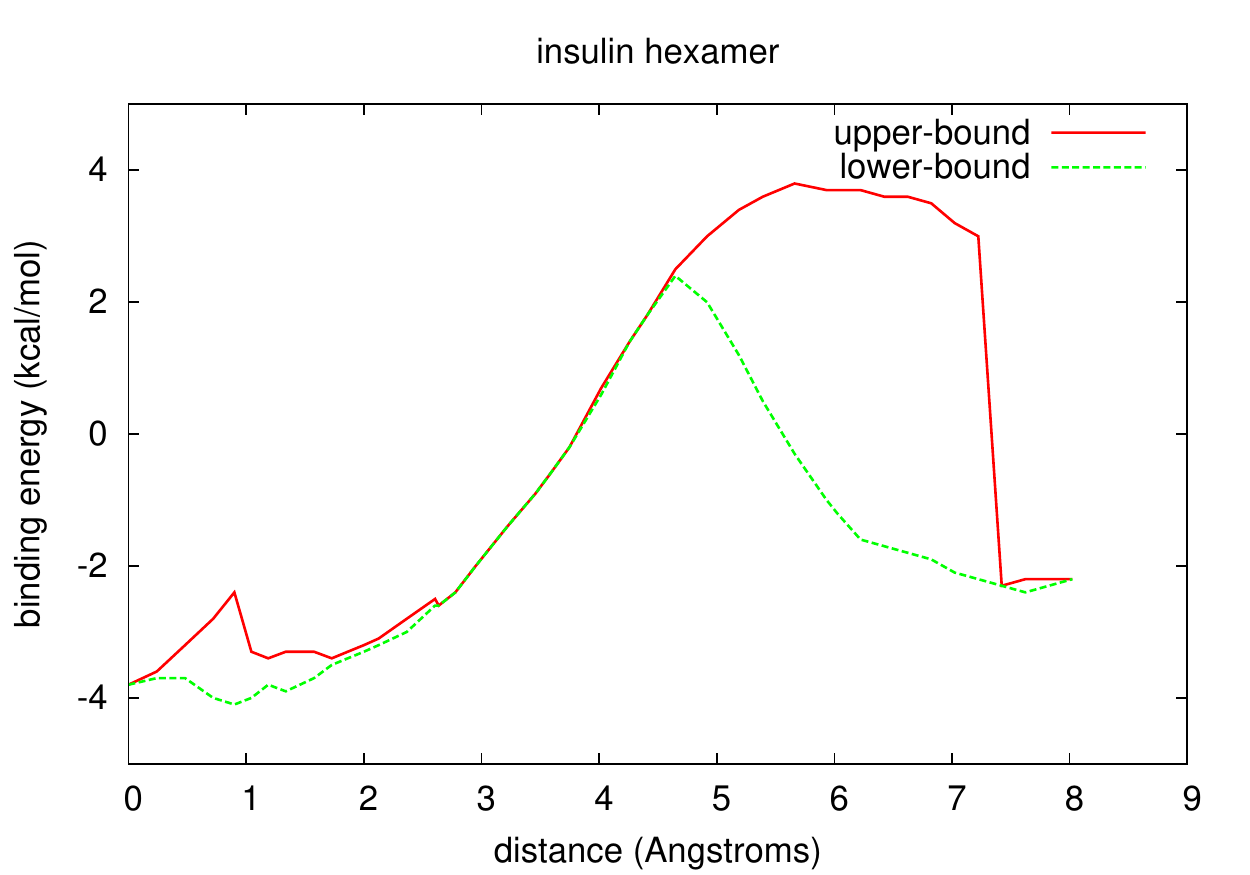}
\includegraphics[width=.32\hsize]{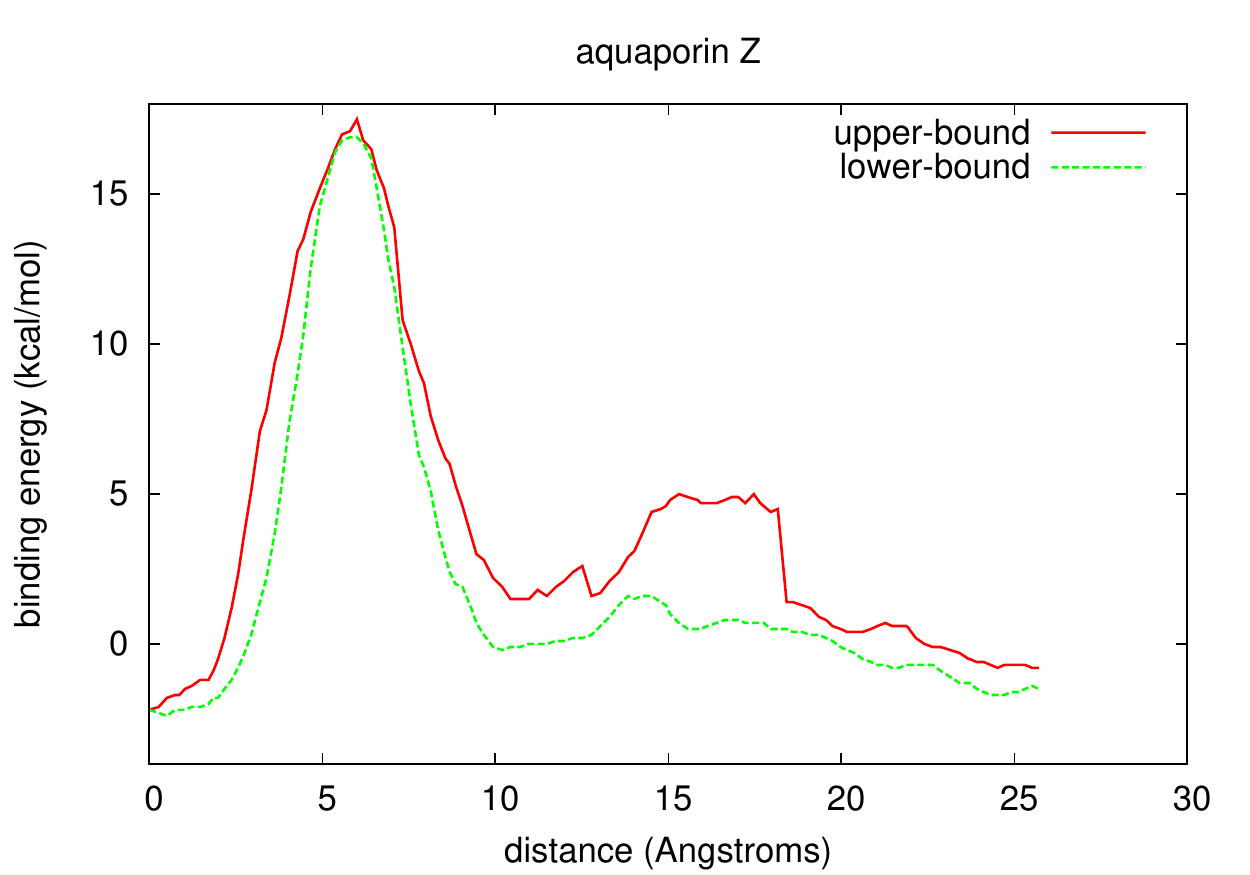}
\includegraphics[width=.32\hsize]{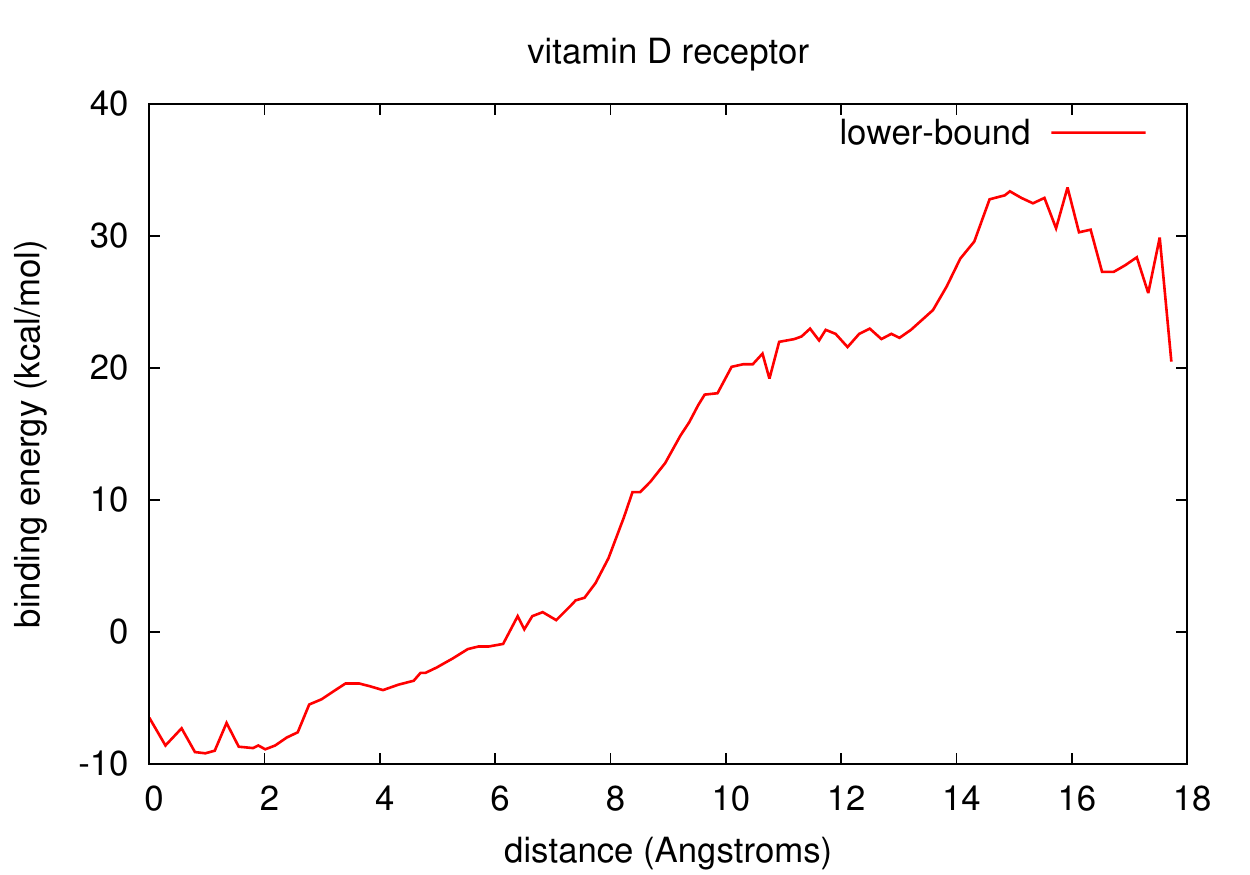}

\includegraphics[width=.32\hsize]{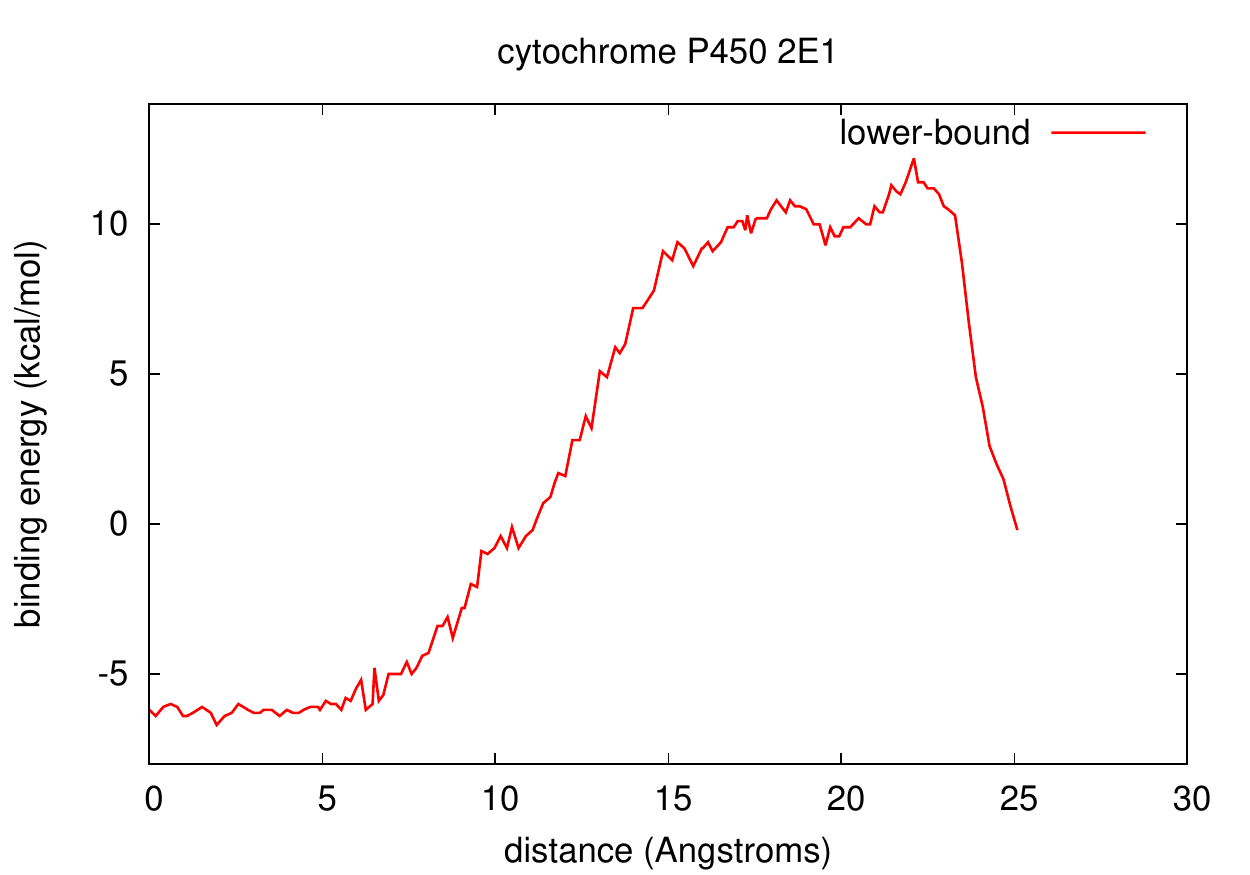}
\caption{Energy profiles of the ligand movements in the test set defined in Table~\ref{tab:complexes}. The distance is measured from the tunnel bottom to the protein surface.}
\label{fig:energies}
\end{figure}

The energy profiles of the tested ligand-protein complexes are shown in Figure~\ref{fig:energies}. They represent the variation of the binding energy of the ligands moved from the active site to the tunnel entrance at the protein surface. 

In all experiments the upper-bound trajectory has higher energy than the lower-bound because the ligand movement is restricted by the pattern constraint, and thus cannot easily overcome small bottlenecks. Whereas the upper-bound energies usually copy the shape of the lower-bound energetic profile, some bottlenecks are visible in upper-bound trajectory only (such as in acetylcholinesterase from distance 7\,\angstrom ~to 16\,\angstrom, in lipase B from 3\,\angstrom ~to 6\,\angstrom ~and aquaporin Z from 13\,\angstrom ~to 18\,\angstrom). As we can see, the contiguous trajectory adds additional information useful for the ligand transport analysis. Note that the observed bottleneck does not necessarily indicate that the transportation of a ligand through a protein tunnel is not possible. The protein flexibility may allow the ligand to pass even higher energy barriers observed in static structures. The interpretation of such data is crucial -- the part of the protein forming the bottleneck may be more or less rigid in a real-world system, or sometimes some form of flexibility may lead to the opening of the tunnel and allow the ligand to pass. The CaverDock user may select the residues to which the side chain flexibility may be introduced. If a protein backbone forms an artificial bottleneck, then a different protein conformation has to be used.

We have compared the CaverDock energy profiles to SLITHER in cases where we were able to compute the SLITHER trajectory. The data provided by SLITHER has been filtered: we have removed all conformations which were not placed within the tunnel determined by CAVER. Such filtering is necessary to exclude any conformations in different tunnels or at the protein surface. The advantage of the constrained docking used in CaverDock can be seen when the energy profiles are compared. The trajectory obtained with SLITHER was sparser when compared to CaverDock and no conformation was placed in the bottleneck. For example, only two conformations were computed in the leucine transporter's tunnel and they are located before and after the bottleneck. In some cases it might be possible to roughly guess the position of the bottlenecks based on the gaps in SLITHER's trajectory. In other cases, SLIGHTER's trajectory may include gaps also at the positions with no observable bottleneck, which can be seen for example in the second half of the trajectory with haloalkane dehalogenase. CaverDock reports the energies and conformations of the ligand in the bottlenecks, so it is possible to analyze, which residues may be mutated to increase the rate of ligand's passage. Note that the absolute energy values are different for SLITHER and CaverDock, which is caused by using different chemical force fields (the constrains force field terms are excluded from CaverDock output energies).

\section{Conclusions and Future Work}
\label{sect:conclusion}

In this paper, we have introduced a novel method for analysis of the transport of ligands in proteins and its implementation in CaverDock tool. We have developed a constrained force field for molecular docking and a heuristics to analyze the ligand movements in the tunnel. We have also introduced a new algorithm for discretization of the protein tunnels. Our approach extends the state-of-the-art by using molecular docking for calculation of contiguous movements of the ligand within the tunnel. The calculation is faster and easier to setup compared to MD but on the other hand overcomes the limitations of geometrical methods. We have demonstrated that CaverDock is robust and is able to analyze the ligand transportation usually in minutes, or in a few hours in the worst scenario.

In the future, we plan to improve CaverDock heuristics to compute more alternative trajectories and to connect promising, non-contiguous parts of trajectories more aggressively. We expect to obtain lower energy for upper-bound trajectories. Or, at least, generate upper-bound trajectories with higher confidence, so the energy is not overestimated due to insufficient sampling. Furthermore, we plan to improve the receptor flexibility in CaverDock. With the current version, only the side-chains can be flexible. The flexibility of the protein backbone would allow to model situations where the receptor's flexibility plays a significant role in the ligand passage. More precisely, we will explore the possibility to use an ensemble of protein conformations or coarse-grained MD to reproduce the movement of the protein backbone.

\section*{Acknowledgments}
The work was supported from Grant Agency of Masaryk University (MUNI/M/1888/2014) and European Regional Development Fund, Pro\-ject "CERIT Scientific Cloud" (No. CZ.02.1.01/0.0/0.0/16\_013/0001802). The authors express their thanks also to infrastructural projects ELIXIR and C4Sys run by the Czech Ministry of Education (LM2015047 and LM2015055) for financial support. Access to the CERIT-SC computing and storage facilities provided by the CERIT-SC Center, under the programme "Projects of Large Research, Development, and Innovations Infrastructures" (CERIT Scientific Cloud LM2015085), is greatly appreciated.

\bibliographystyle{plain}
\bibliography{fila}

\end{document}